\documentclass{article}
\usepackage[a4paper, total={6.5in, 10in}]{geometry}
\usepackage[utf8]{inputenc} 
\usepackage[T1]{fontenc}
\usepackage[hidelinks]{hyperref}       
\usepackage{url}            
\usepackage{booktabs}       
\usepackage{amsfonts}       
\usepackage{nicefrac}       
\usepackage{microtype}      
\usepackage{lipsum}
\usepackage{fancyhdr}       
\usepackage{graphicx}       
\graphicspath{{media/}}     
\usepackage{amsmath}        

\usepackage[labelfont=bf]{caption}

\usepackage{graphicx}
\usepackage{adjustbox}
\usepackage{subcaption}

\usepackage{mathtools}

\usepackage{bm} 
\usepackage{bbm} 
\usepackage{dsfont} 

\usepackage{algorithm}
\usepackage[noend]{algpseudocode}

\usepackage{xcolor}

\usepackage{comment} 

\usepackage{sidecap}

\usepackage{booktabs}

\usepackage{cite} 
\usepackage{lscape}
\usepackage{authblk}
\usepackage{tabularray}

\title{Measuring the co-evolution of online engagement with (mis)information and its visibility at scale}
\date{}

\author[1,2,3]{Yueting Han}
\author[4]{Paolo Turrini}
\author[2,5]{Marya Bazzi}
\author[6,7,8]{Giulia Andrighetto}
\author[6]{Eugenia Polizzi}
\author[9,10,11]{Manlio De Domenico}

\affil[1]{\small MathSys CDT, University of Warwick, Coventry, UK}
\affil[2]{\small Mathematics Institute, University of Warwick, Coventry, UK}
\affil[3]{\small The Alan Turing Institute, London, UK}
\affil[4]{\small Department of Computer Science, University of Warwick, Coventry, UK}
\affil[5]{\small sea.dev, London, UK}
\affil[6]{\small Institute of Cognitive Sciences and Technologies, National Research Council of Italy, Italy}
\affil[7]{\small Institute for Futures Studies, Stockholm, Sweden}
\affil[8]{\small Institute for Analytical Sociology, Linköping University, Sweden}
\affil[9]{\small Department of Physics and Astronomy ‘Galileo Galilei’, University of Padua, Padova, Italy}
\affil[10]{\small Padua Center for Network Medicine, University of Padua, Padova, Italy}
\affil[11]{\small Istituto Nazionale di Fisica Nucleare, Sez. Padova, Italy}

\begin{document}

\newcommand{\stanceA}{factual}
\newcommand{\stanceB}{misleading}

\maketitle
\begin{abstract}

Online attention is an increasingly valuable resource in the digital age, with extraordinary events such as the COVID-19 pandemic fuelling fierce competition around it. 
As misinformation pervades online platforms, users seek credible sources, while news outlets compete to attract and retain their attention. 
Here we measure the co-evolution of online ``engagement'' with (mis)information and its ``visibility'', where engagement corresponds to user interactions on social media, and visibility to fluctuations in user follower counts.
Using over 100 million COVID-related retweets across 3 years, we analyse how user interactions and follower dynamics differ for factual, misleading and uncertain content. 
We observe that during major events (e.g., vaccine rollouts), users spreading factual content see rapid follower gain spikes, whereas those sharing misleading content tend to sustain faster growth outside of these high-attention periods. 
We introduce two scalable modelling frameworks (simple contagion and biased convergence) that reproduce many observed differing follower growth rates using temporal retweet network dynamics, providing evidence that content visibility co-evolves with user engagement.
Our modelling lends itself to studying other large-scale events where online attention is at stake, such as climate and political debates.

\end{abstract}

\section{Introduction} 
False, misleading, or unreliable content in social media has been a long-standing concern on various topics, e.g., politics \cite{Flamino-2023, Bovet-2019, Pennycook-2018}, environment \cite{Spampatti-2023, Biddlestone-2022, Farrell-2019}, health \cite{Allen-2024, Loomba-2021}.
As the public becomes increasingly dependent on social media for information, it plays an important role in shaping public opinion, with misinformation being an ever-present and increasing threat. 
This ties to the phenomenon of ``fight for attention'', where different narratives aim to increase their content visibility, as has been studied in \cite{Trielli-2019, Giraldo-Luque-2020, Feng-2015}.
Such narratives need to be grounded in scientific evidence and we often find questionable sources being more successful than scientifically solid ones. 
But what makes a source win the battle for users and does misinformation spread in a way that is coupled to scientific information? 

Substantial research has investigated misinformation on social media \textemdash its origins \cite{Farrell-2019, Ferrara-2016}, detection \cite{Abdali-2024, Shu-2017}, flow \cite{Goel-2025, Allen-2024, Flamino-2023, Prollochs-2023, Bovet-2019, Vosoughi-2018, Vicario-2016}, and sentiments \cite{Liu-2024, Farhoundinia-2024, Prollochs-2021}.
Public opinion, on the other hand, is often measured through controlled experiments, in which participants are exposed to curated misinformation or factual content on a topic and their responses are evaluated through surveys \cite{Allen-2024, Spampatti-2023, Loomba-2021, Pennycook-2018}.
Recent studies focus on integrating these measurements of online (mis)information diffusion and public opinion to track, in real time, how belief dynamics evolve at scale as information spreads across networked platforms \cite{Aslett-2024, Tokita-2024, Guess-2019}.
For example, Tokita et al. \cite{Tokita-2024} conduct a two-fold investigation on top-trending Twitter articles (102 true, 37 false or misleading through manual verification) \textemdash for each article, they use observational Twitter data (i.e., the followers of users who shared it) to estimate user exposure and deploy real-time surveys, conducted at the time of each article's release, to approximate the likelihood that exposed users believe the content is true.

Inspired by these recent studies, which often employ a dual approach to measure (mis)information and opinion diffusion jointly, this paper proposes a unified modelling framework that integrates both through follower count dynamics.
Follower count measures the number of people subscribed to a user's account, enabling them to receive notifications to view that user's activities.
As such, for spreaders of \stanceA{} or \stanceB{} content, tracking their follower count serves both as a useful metric for estimating their audience and, through its dynamics over time, as an indicator of social endorsement, and possibly opinion shifts \textemdash with users following content spreaders acting as digital footprints of ``belief'', or at minimum, deliberate engagement.
Here, we associate this dynamic with interactions among users that signify positive information flow (e.g., repost, recommendation), which, as noted in various studies, prompts a user's followers to extend to others, thereby facilitating follower growth \cite{Myers-2014, Johnson-2020, Han-2024, Hutto-2013}.
However, this interaction-follower gain dynamic is highly heterogeneous across individuals, as changes in follower count are influenced by factors beyond just interactions \cite{Myers-2014, Bertone-2022, Mueller-2017, Garcia-Martin-2016, Metaxas-2021}. 
For instance, Myers \& Leskovec \cite{Myers-2014} find on Twitter that retweets and follower gains often occur in sequential hourly bursts after a tweet's release, while an unfollow burst can also happen, possibly due to tweets of offensive content that disengage followers.
Another study by Bertone et al. \cite{Bertone-2022} reveals that, on a long-term scale of years, the follower count of celebrities (e.g., musicians, athletes) on Instagram is influenced by trending topics, which can be measured externally through tools such as Google Trends for search volume.

In light of this inherently noisy interplay between interactions and follower gain at the individual level, we coarse-grain our modelling toward group dynamics involving (mis)information on a specific topic of interest.
Aside from segmenting groups based on topic credibility (\stanceA{} or \stanceB{}), as considered in many previous studies \cite{Cinelli-2021, Vicario-2016, Sikder-2020, mcpherson2001birds}, this paper includes an additional group of users who engage with content related to the topic, but the majority of such content cannot be easily categorised as mostly veering towards ``\stanceA{}'' or ``\stanceB{}'' (referred to as ``uncertain''), e.g., satire or material that, while related to the topic, reflects interests in other areas. 
This group serves as a benchmark for comparing the behaviour patterns of the other two groups, i.e., users primarily spreading \stanceA{} or \stanceB{} content within the topic.
We assume that users in those groups gain followers by spreading information to the remaining users \textemdash those who engage with a mix of \stanceA{}, \stanceB{} or uncertain content, thereby swaying their followers to join their campaign.
The mechanism behind the group dynamic we investigate builds upon prior work \cite{Johnson-2020, Han-2024, Nicola-2023}, with \cite{Johnson-2020, Han-2024} simulating follower count growth in a dataset with limited timestamps and size.
Specifically, they analyse a Facebook dataset that describes two snapshots in 2019 of online recommendations among 1326 pages on the topic of vaccines, where each page is assigned a vaccine stance (anti-, pro-, or undecided/neutral) through manual check.
They simulate information spreading from anti- and pro- to neutral groups, through which they predict follower gains for anti- and pro- groups, using models such as an ODE (ordinary differential equation) system \cite{Johnson-2020} and the epidemic framework \cite{Han-2024}.

We apply our modelling frameworks built on \cite{Johnson-2020, Han-2024}, which explores how interactions co-evolve with follower gain, to 
a unique Twitter\footnote{Although Twitter has been rebranded as ``X'', we use the name ``Twitter'' in this paper, as it was the name in use at the time of data collection.} dataset of over 100 million retweets related to COVID-19, collected from 17 March 2020 to 12 February 2023, covering the early to late stages of the pandemic.
Each retweet is classified as \stanceA{}, \stanceB{}, or uncertain based on the web links it contains (if any), using a well-established database of nearly 4,000 expert-curated domains assessed for scientific credibility.
This database was originally developed and applied in Gallotti et al. \cite{Gallotti-2020} and later used in Castioni et al. \cite{Castioni-2022} to explore social dynamics in misinformation circulation at the early stage of the dataset.
In this paper, with a dataset extending over a three-year timespan, we track the follower count fluctuations of users involved in retweeting or being retweeted, covering around 14 million users and totalling approximately 30 billion followers.
We also collect their Twitter verification status and bot detection to provide additional insights.
Due to the large size of our dataset, some necessary pre-processing is performed to reduce computational complexity while preserving key properties.

\paragraph{Contribution}
In contrast to previous research where typically (mis)information and belief diffusion are assessed independently, this paper proposes a unified modelling framework that measures both through the follower gains driven by user interactions on social media. 
This offers a more accessible alternative to resource-intensive survey data, and a more scalable solution via network-based statistical filtering.
We expand the framework upon earlier studies \cite{Johnson-2020, Han-2024}, which use a dataset with limited timestamps and size.
Our investigation here is anchored in a large-scale Twitter dataset of retweets related to the COVID-19 pandemic, spanning three years.

Our contribution is three-fold. 
First, we scale down the dataset while preserving key users likely to gain followers through retweets, along with other essential properties, using a network backbone method known as disparity filter \cite{Serrano-2009}.
Second, we track follower count fluctuations of news sources with varying scientific credibility, finding that while they generally grow, the trends of those primarily spreading either \stanceA{} or \stanceB{} content differ in a way that is tightly and intuitively coupled to external factors\textemdash deeply tied to the COVID-related events (e.g., vaccine roll-out, epidemic severity), whose early-stage patterns are consistent with \cite{Gallotti-2020, Castioni-2022} where alternative measures are used.
Third, building on \cite{Johnson-2020, Han-2024}, we propose two frameworks (i.e., simple contagion and biased convergence) that reproduce many of the observed differing follower growth rates using temporal retweet network dynamics, providing evidence that follower growth co-evolves with retweet dynamics.
Our modelling frameworks are general and could serve as a basis for linking the follower growth dynamics among users beyond the types of interactions and topic investigated in this study (here, retweets involving \stanceA{}, \stanceB{}, or uncertain content related to COVID-19).

\section{Data description}
The dataset contains around 113.7 million retweets \textemdash interactions that allow users to rebroadcast the post (i.e., tweet) of another user to their followers on Twitter, collected second-by-second from 17 March 2020 to 12 February 2023, across countries with all content in English.
These retweets are identified as being related to COVID-19, and each is classified into one of three types: \stanceA{} (36.9\%), \stanceB{} (7.7\%) and uncertain (55.4\%). 
Approximately 13.9 million users are involved, either retweeting or being retweeted, with an estimated total follower count of 28.3 billion at the start (17 March 2020) and 32.4 billion at the end (12 February 2023) of the dataset.

The raw data is collected by the COVID-19 Infodemic Observatory (\url{https://covid19obs.fbk.eu}), following the approaches established in \cite{Gallotti-2020} and later applied in \cite{Castioni-2022}.
These studies cover the early stages of the dataset used here: three months (January - March 2020) in \cite{Gallotti-2020} and five months (January - May 2020) in \cite{Castioni-2022}.
Below, we outline the data collection approaches relevant to this paper, along with the data processing steps performed in this study to support the modelling.

\begin{itemize}
    \item \textbf{Retweet}: 
    Using the Twitter API (application programming interface), public retweets related to COVID-19 are collected second by second during the observation period.
    These retweets are identified when either the original tweet or the content added during retweeting contains any of the following hashtags or keywords: ``coronavirus'', ``ncov'', ``\#Wuhan'', ``covid19'', ``covid-19'', ``sarscov2'', ``covid''.
    The chosen keywords are not intended to comprehensively capture all COVID-related retweets, but to include the most persistent and widely used terms that remain in use even amid abrupt shifts in discourse (e.g., variants such as Omicron, or lockdowns).
    Note that the Twitter API, by default, caps the number of retrievable tweets that meet our selection criteria at 1\% of the total volume of tweets posted per second.

    These COVID-related retweets are then classified into \stanceA{}, \stanceB{}, or uncertain based on the URLs they contain (if any), which direct users who read the retweets to corresponding websites.
    To achieve this, the same manually checked web domains database is used as in \cite{Gallotti-2020, Castioni-2022}, carrying over their assessments of scientific credibility.
    The database integrates nine publicly available data sources, screening a total of 3920 expert-curated domains, with potential biases at different levels (e.g., from the Twitter filtering API or manual annotations) carefully addressed (see details in \cite{Gallotti-2020}).
    Each domain is classified by expert annotators into one of ten types reflecting varying levels of scientific credibility: ``Science'', ``Mainstream media'', ``Satire'', ``Clickbait'', ``Political'', ``Fake or hoax'', ``Conspiracy and junk science'', ``Other'', ``Shadow'', ``NA''. See Table 1 of \cite{Gallotti-2020} for category descriptions; here ``NA'' refers to around 26.0\% of those with URLs but not included in our URLs database.
    Each such retweet is then assigned the category of the corresponding domain the URLs lead to, classifying it into one of the ten types.
    For the purpose of this paper, we group categories of ``Science'', ``Mainstream media'' as \stanceA{}, categories of ``Fake or hoax'', ``Conspiracy and junk science'', ``Clickbait'' as \stanceB{} \textemdash consistent with \cite{Castioni-2022}, and the remaining categories (i.e., ``Satire'', ``Political'',  ``Other'', ``Shadow'', ``NA'') as uncertain. 
    
    \item \textbf{Users - retweeters \& retweetees}:
    For each COVID-related retweet collected and categorised above, both the users who retweeted (retweeters) and those who were retweeted (retweetees) are recorded in the dataset.
    
    Whenever a user is recorded retweeting or being retweeted, the dataset logs their follower counts at the moment of the activity.
    This allows for close monitoring of users' follower count variations during active engagement with COVID-related content, while disregarding fluctuations during long periods of inactivity \textemdash these are not of interest in our paper, as they are potentially irrelevant to the COVID-related retweets. 
    We come back to this point when discussing our results.

    To offer additional insights, the dataset also records whether users are verified by Twitter and whether they are detected as bots (i.e., automated online agents) through the machine learning method \cite{Stella-2019}, each time they are recorded retweeting or being retweeted. 
\end{itemize}

\section{Results}
\subsection{Extracting backbone of retweet network} 
Using the COVID-related retweets dataset described in \S2, we start by constructing a single directed weighted network, referred to as the ``retweet network'', that aggregates retweets over the entire dataset timeframe, from 17 March 2020 to 12 February 2023.
When a user $j$ retweets a user $i$ during this period for $w_{ij} > 0$ times, we add an edge from node $i$ to node $j$ with weight $w_{ij}$ in this network. 
The direction of the edges represents the direction of the information flow (i.e., a tweet from user $i$ capturing the interest of user $j$, leading to user $j$ retweeting user $i$). 
The resulting retweet network contains around 13.9 million nodes and 72.3 million edges, with the highest edge weight reaching 23.2 thousand. 

To reduce the computational complexity in subsequent modelling steps, and to focus on influential users who are more likely to gain followers from being retweeted for COVID-related content, we reduce the size of our retweet network by applying a widely used network backbone extraction method known as disparity filter \cite{Serrano-2009}.
This method retains only the edges with statistically significant weights at a local scale, as compared to a null model.
Loosely speaking, we keep an edge from user $i$ to $j$ if it falls into either of these two cases below: user $j$ retweets user $i$ a lot more than they retweet others; conversely, user $i$ is retweeted by user $j$ much more than by others. 
In either case, user $i$ has the potential to influence user $j$ and attract followers from user $j$.
See further details about this method in \S5.1.

\begin{figure}[ht]
    \centering
    \adjincludegraphics[width=.95\textwidth,trim={{.00\width} {.00\height} {.00\width} {.00\height}},clip]{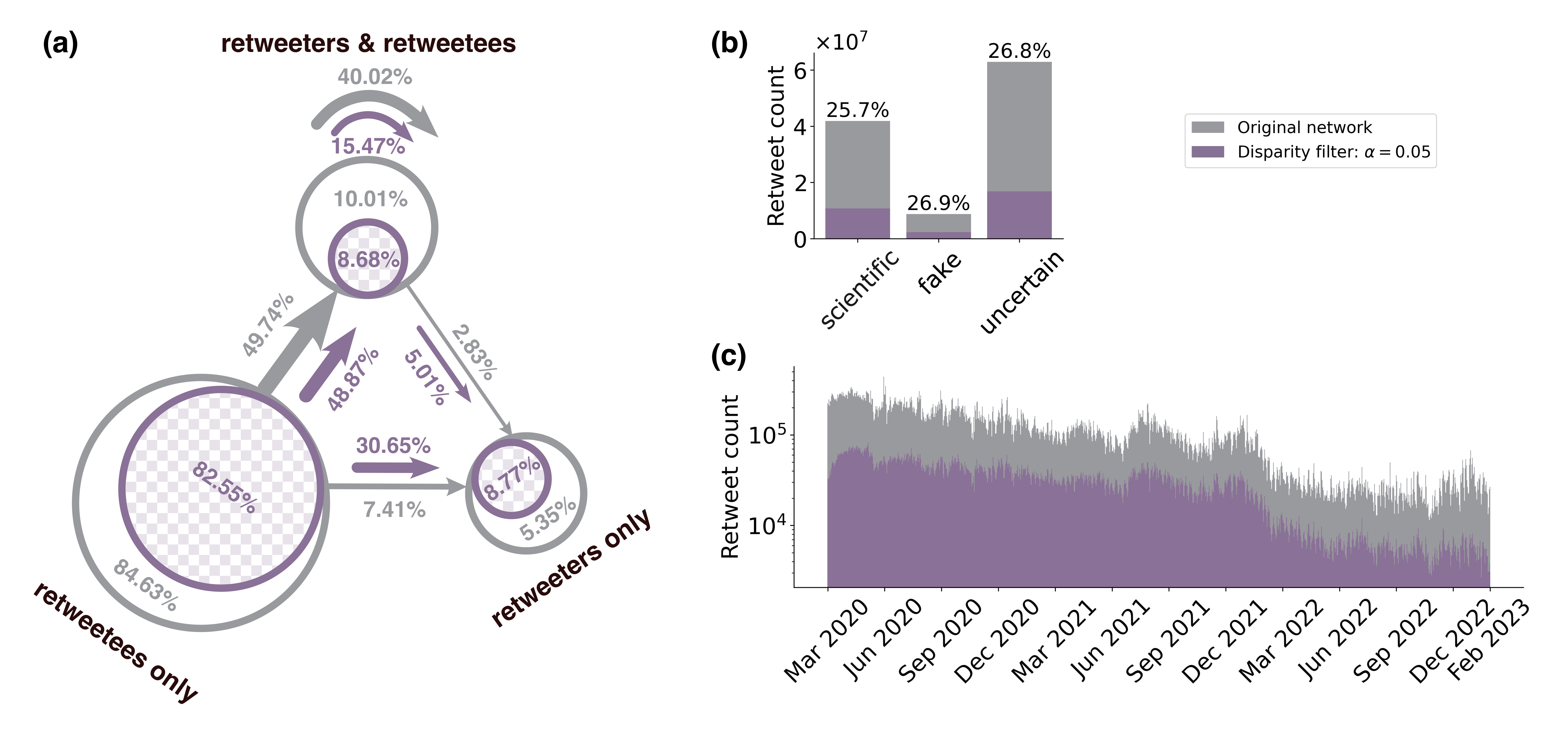}
    \caption{\textbf{Comparison of the original and filtered retweet network.} The filtered retweet network (purple) preserves key features of the original one (grey) across the entire timeframe from 17 March 2020 to 12 February 2023.
    \textbf{(a) Retweeter-retweetee dynamics.} ``retweetees only'' includes users who are only retweeted, ``retweeters only'' includes those who only retweet, and ``retweeters \& retweetees'' includes the rest.
    Each component's percentage indicates the proportion of users it contains, and each edge's percentage represents the proportion of retweets between two components. 
    Most percentages associated with the original network are similar to those in the filtered network.
    \textbf{(b) Retweet category distribution.} The percentage on each bar shows how much of the retweets from that specific category (i.e., \stanceA{}, \stanceB{}, or uncertain) in the original network are retained in the filtered network. 
    These percentages consistently remain around 25\% across all three categories.
    \textbf{(c) Retweet temporal distribution.} The filtered network preserves the temporal distribution shape in comparison to the original network, capturing both the long-term decline trend and short-term spikes seen in the original network.
    }
    \label{fig:netbone}
\end{figure}

We filter the network at the significance level $\alpha = 5\%$, retaining approximately 0.8 million nodes (5.5\%), 2.2 million edges (3.0\%), and 30.0 million weight (26.4\%). 
Figure \ref{fig:netbone} shows that key features are preserved compared to the original network.
On the one hand, the filtered network exhibits a highly asymmetric 
``retweeter-retweetee''
pattern, similar to the original network (Figure \ref{fig:netbone}a) \textemdash over 80\% users acting solely as retweetees and less than 10\% solely as retweeters.
Interestingly, among the remaining only 10\% of users both retweeting and being retweeted, the largest strongly connected component consistently includes far more users than the second largest in both the original network (1st: 777,152; 2nd: 110) and the filtered network (1st: 10,562; 2nd: 30).
On the other hand, although we disregard retweet categories (\stanceA{}, \stanceB{}, uncertain) and timestamps during filtering, interestingly their distribution shapes are also preserved.
Around 25\% of retweets across each content type are consistently retained in the filtered network (Figure \ref{fig:netbone}b). 
The temporal distribution of retweets in the filtered network also captures both the long-term decline trend and short-term spikes seen in the original network (Figure \ref{fig:netbone}c).

We discuss the robustness of our results below, with details available in Supplementary Information A.
(1) The disparity filter assumes a system with strong disorder, where weights are heterogeneously distributed both globally and locally \textemdash a criterion that our dataset satisfies.
(2) The choice of significance level $\alpha = 5\%$ is made, so as to keep the filtered network's topological properties (i.e., degree distribution, weight distribution, clustering coefficient) close to the original network, while minimising the network's size.
(3) We find that more bot users are filtered out compared to non-bot users, and this observation is supported by the analysis of users' verification status on Twitter, where more verified users are retained.
(4) When further examining retweet categories that are not generalised (as introduced in \S2), we observe varying retention percentages, which are notable but unsurprising: ``CONSPIRACY/JUNKSCI'' (\stanceB{}) and ``SATIRE'' (uncertain) exhibit the highest retention percentages (40.0\% and 34.0\%, respectively), whereas ``OTHER'' (uncertain) shows the lowest retention percentages (21.6\%). 

Unless stated otherwise, all results discussed in the following sections are based on the filtered retweet network.

\subsection{Retweet vs follower count variations by category}
From the filtered retweet network, we identify users from different campaigns (\stanceA{}, \stanceB{}, uncertain) based on the retweet categories they engage with. 
This allows us to track changes in follower counts for each campaign over time, in contrast to their retweet counts. 

\begin{figure}[ht]
    \centering
    \adjincludegraphics[width=0.85\textwidth,trim={{.0\width} {.0\height} {.0\width} {.0\height}},clip]{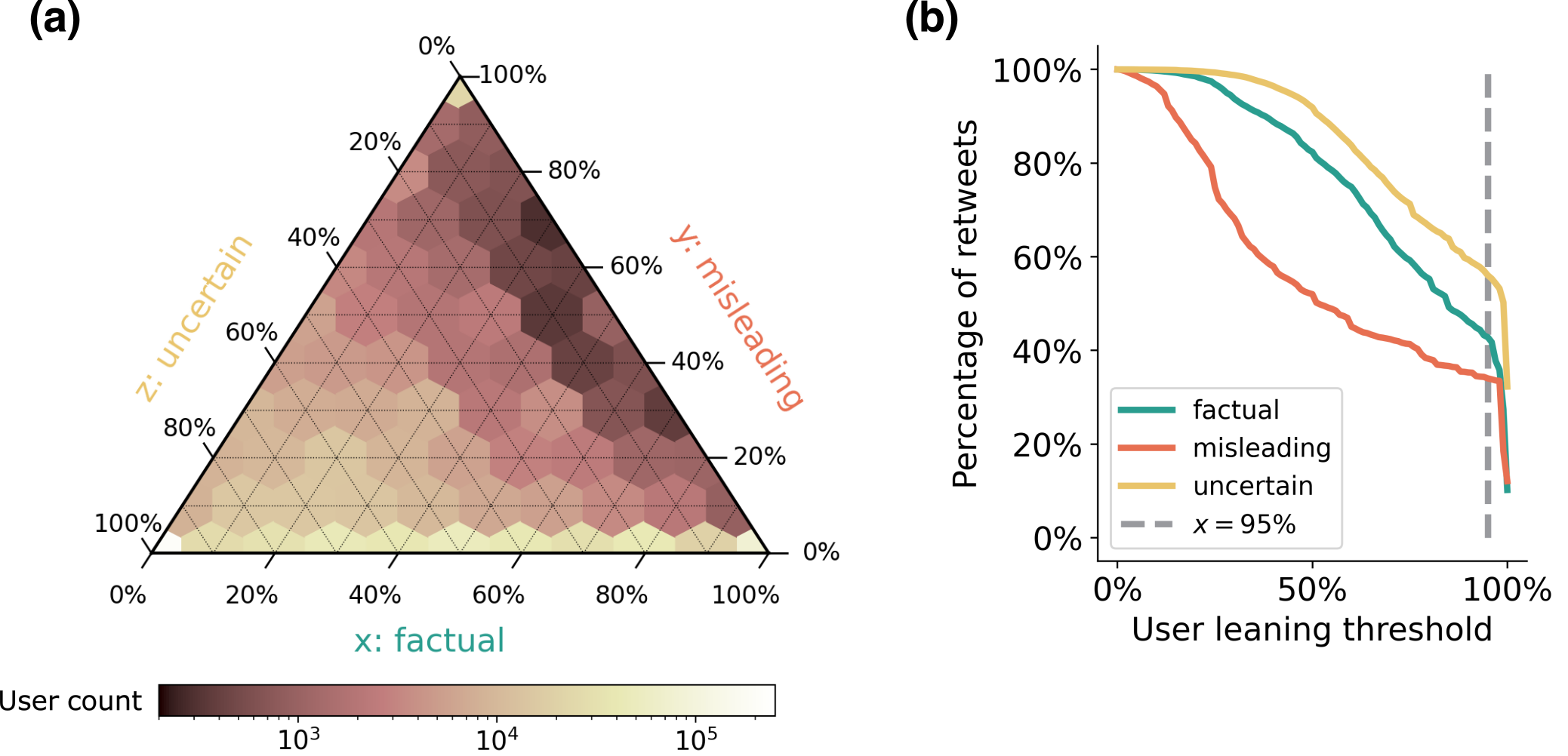}
    \caption{\textbf{Thresholding highly aligned users.} Using the filtered retweet network, we categorise users as highly aligned with a campaign (\stanceA{}, \stanceB{}, or uncertain) if >95\% of the retweets they give or receive are of that content type. 
    \textbf{(a) User distribution by retweet proportions for different content types.} The triangle heatmap depicts the user distribution based on the proportion of retweets for each content type that every individual user gives or receives. Three light-coloured corners (>95\%) suggest a large number of users predominantly circulate only one type of content. Darker cells along the y-axis suggest users are much less inclined to circulate \stanceA{} and \stanceB{} content together. 
    \textbf{(b) Percentage of retweets by highly aligned users at varying thresholds. } For each content type, we calculate the percentage of corresponding retweets given or received by highly aligned users at varying thresholds. The figure displays notable jumps around the 95\% threshold for all three content types, despite some differences in curve shapes. This indicates that users highly aligned (>95\%) are involved in a large proportion of retweets.} 
    \label{fig:obs1}
\end{figure}

For users engaged in diverse content categories, it can be challenging to determine which one helps gain followers.
Therefore, in the filtered network, we analyse the proportion of retweets each user involves (either gives or receives) across these categories, to identify users who are highly aligned with each specific content category\footnote{Note that, while this paper focuses on the dynamics of users gaining followers through receiving retweets, including both given and received retweets here allows us to also categorise users who only give retweets.
In fact, we find that for users who both give and receive retweets, the content types they give largely match those they receive, leading to primarily consistent results, as shown in Figure \ref{fig:obs1}.}.
Figure \ref{fig:obs1}a indicates that a large number of users primarily engage with a single content category. Users also rarely engage with both \stanceA{} and \stanceB{} content, reflecting real-world dynamics.
As a result, we consider a user to be highly aligned with a campaign if >95\% retweets they involve are of that content type\footnote{A user retained in the filtered network involves in $\geq$4 retweets during the data timeframe.}. This classifies about 73.0 thousand users as \stanceA{}, 19.0 thousand as \stanceB{}, and 208.8 thousand as uncertain, totalling 39.2\% of all users. 
This threshold choice is supported when we investigate, for each content type, the percentage of retweets circulated by highly aligned users at varying thresholds (Figure \ref{fig:obs1}b). The notable jumps for all three content types around the 95\% threshold suggest that highly aligned users at this threshold are involved in a large proportion of retweets.
When examining the profiles of these highly aligned users, we find those most engaged with \stanceA{} content are primarily from mainstream media. 
In contrast, users of \stanceB{} content are often daily news outlets, possibly aiming to attract public attention and increase online traffic. 
Some users of uncertain content focus on politics, religion, or business, suggesting that COVID may not be their primary concern.

\begin{SCfigure}[][htbp]
    \centering
    \adjincludegraphics[width=0.65\textwidth,trim={{.00\width} {.00\height} {.00\width} {.00\height}},clip]{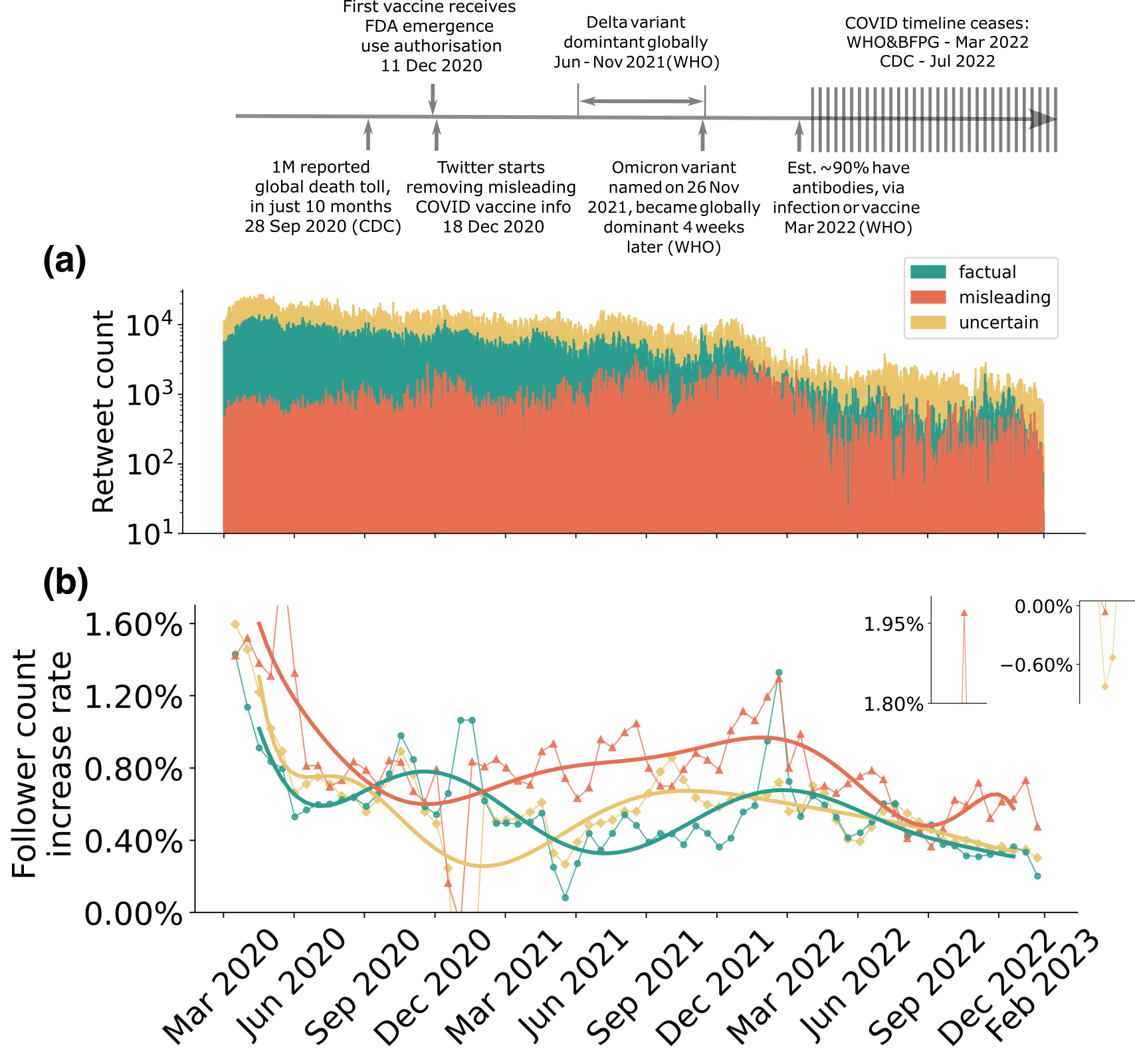}
    \caption{\small \textbf{(a) Retweet vs (b) follower count variations by category.} The COVID-related events are obtained from \cite{CDC, WHO, WHO1, WHO2, BFPG, Twitter}.
    \textbf{(a)} On a daily basis, we collect the count of retweets given or received by users who are highly aligned with each campaign.
    We visualise three overlapping temporal distributions by content type, with uncertain content consistently having the highest retweet volume, followed by \stanceA{} and then \stanceB{} content.
    \textbf{(b)} Each data point of follower increase rate is obtained using a one-month time window, which is shifted forward by half a month throughout the data timeframe. 
    In every window, we select from the pool of highly aligned users the active ones who retweet or are retweeted at least twice (to ensure their follower counts are recorded at least twice). 
    To obtain the follower count increase rate for each campaign, we compare the aggregated follower counts of active aligned users recorded at the first and last log within the window.
    Trend lines are generated using 5-point boxcar smoothing, followed by 10-degree polynomial regression.
    The upper right subfigures show areas exceeding the main figure's range around May 2020 and January 2021.
    }
    \label{fig:obs2}
\end{SCfigure}

Using the classified highly aligned users, we then track the variations in follower counts for each campaign. 
Recall that, as mentioned in \S2, our dataset logs follower counts only when users retweet or are retweeted. This captures follower changes during active engagement with COVID-related content, while excluding fluctuations likely unrelated to COVID during periods of inactivity.
Given that the average activity gap for highly aligned users is around 19 days, we adopt one-month time windows to track follower count changes, focusing on users active at least twice within each window.
Specifically, for each campaign within a time window, we compare the total follower counts of active highly aligned users at the start and end of the window to obtain the increase rate. 
We slide the time window every half a month to add more data points.

The results are shown in Figure \ref{fig:obs2}, along with the daily retweet counts for each campaign as a contrast.
At the start of the dataset, the total follower counts of users aligned with \stanceA{}, \stanceB{}, and uncertain content are approximately 929.9 million, 43.6 million, and 3.0 billion, respectively (see Supplementary Information B for the distribution of individual-level follower counts).
As for follower count fluctuations over time, Figure \ref{fig:obs2} illustrates an overall growth trend across all content types, with the largest increase around the initial COVID-19 outbreak in early 2020.
This was followed by a gradual slowdown in the growth rate, particularly after mid-2022, suggesting a possible waning interest in COVID-related topics, or that hashtags and keywords we use to track COVID-related tweets no longer capture mainstream discussions effectively in the later stage.

Some differences in follower count growth trends exist across various content types in Figure \ref{fig:obs2}.
For users primarily engaged with \stanceA{} content, their follower count shows several local peaks, notably three occurring right after: 
(1) the announcement that the global COVID-19 death toll surpassed 1 million in 10 months \cite{CDC};
(2) the FDA (U.S. Food and Drug Administration) granted EUA (Emergency Use Authorisation) for the first COVID-19 vaccine by Pfizer-BioNTech \cite{CDC};
(3) the rapid shift from Delta to Omicron variant, with Omicron being more transmissible but causing milder symptoms than Delta \cite{WHO, WHO2}. Note that Delta and Omicron are two prominent variants that stood out among other variants, dominating during specific periods \textemdash see \cite{WHO2, Klamser-2023} for a comparison with other variants.
In contrast, users aligned with \stanceB{} content, while briefly surpassed by those aligned with \stanceA{} during their peak periods, show a higher overall follower growth rate outside these peaks.
Notably, the significant slowdown in follower growth for \stanceB{} content in late 2020 was likely triggered by Twitter's efforts to remove misleading information about COVID-19 vaccines \cite{Twitter}. This was followed by a rapid rebound, peaking in early 2022 when the Omicron variant was dominant.
The follower count for users aligned with uncertain content, on the other hand, experiences relatively mild fluctuations. 
Lastly, we find retweet count trends correlated with variations in follower counts overall, especially for \stanceA{} and \stanceB{} content. Retweet counts peak, with follower count fluctuations often lagging slightly behind, implying potential causality.

Our findings are consistent with and complementary to those of previous studies \cite{Gallotti-2020, Castioni-2022}, which examine the early stage of this dataset from alternative perspectives.
Gallotti et al. \cite{Gallotti-2020} observe a rise of credible sources as infections increased, albeit with high variations across countries \textemdash a pattern mirrored in the follower growth rate of \stanceA{} content, which peaks and shows the most pronounced advantage over \stanceB{} content when the pandemic was most severe.
Castioni et al. \cite{Castioni-2022} find that most fake news was created by a minority of users but consumed by a majority, a pattern also observed in our dataset (see Supplementary Information C).
This potentially explains the relatively higher follower growth rate of \stanceB{} news over the \stanceA{} throughout much of the dataset timeframe.

\subsection{Simulating follower count growth using temporal retweet networks}


Here we explore the coupling dynamics between retweet and follower growth.
That is, whether the evolving retweet dynamics in Figure \ref{fig:obs2}a (a proxy for information flow) can reproduce the temporal variations in follower counts across content types (a proxy for online attention) in Figure \ref{fig:obs2}b, as evidence that attention co-evolves with information flow.

We establish the following three assumptions for our simulations to centralise our research focus and address some limitations in our dataset records.
(1) We model only follower gains without accounting for decreases, assuming that retweets indicate positive information flow that contributes to follower growth, as evidenced by previous studies \cite{Myers-2014, Hutto-2013, Metaxas-2021}.
This is also the case in Figure \ref{fig:obs2}b, where follower count shows an overall growth trend.
(2) Given that our dataset logs follower counts only at the time of retweet activity (average activity gap of around 19 days, potentially delaying change detection), this limits us from conducting causal analysis between retweets and follower growth dynamics.
Instead, we explore their coupling on a coarse monthly timescale, whether retweet networks aggregated from the past $n$ months (referred to as ``temporal retweet networks''), are coupled to follower count growth in the subsequent month.
We consider these temporal retweet networks as revealing potential pathways of positive information diffusion, with follower gains driven by information cascades along these pathways.
By varying the value of $n$, we explore whether follower count growth is coupled to the long-term accumulation of information-flow pathways, or rather sensitive to short-term changes in these pathways.
(3) For our interest in this paper, recall from the introduction that we model group dynamics, specifically targeting aggregated follower count growth for each group of highly aligned users (spreading \stanceA{}, \stanceB{}, or uncertain content).
We adopt a mechanism built on \cite{Johnson-2020, Han-2024}, where highly aligned user groups gain followers by cascading information to ``swayable users'', who engage with a mix of content types and influence their followers to join their campaigns.

Under the assumptions above, we explore two frameworks, ``simple contagion'' and ``biased convergence'', each simulated to capture the co-evolving dynamics of retweets and followers.
We base the simple contagion framework on the classic epidemic SIR (susceptible–infected–recovered) model \cite{Matt-2008}, motivated by the intuition that information spreading via retweets is epidemic-like, with followers gained as users become ``infected''.
We base the biased convergence framework on the classic FJ (Friedkin–Johnsen) opinion dynamics model, motivated by the intuition that retweets act as social influence, with followers gained as their opinions are updated toward more extreme values.
The former yields better results than the latter in aligning with the empirical follower growth data, while the latter still exhibits similarities to the former that support our findings below.
We also learn from the latter (which uses a more relaxed setup without calibration) that certain calibrations are necessary, and we then incorporate them into the former.
For brevity, we present below only the former, i.e., the simple contagion framework, while the latter is detailed in Supplementary Information E.

We first introduce the classic epidemic SIR model, on which our simple contagion framework later builds.
Originally developed to simulate disease spread \cite{Matt-2008}, the SIR model has been adapted to characterise information diffusion on social media \cite{Cinelli-2021, Cinelli-2020, Han-2024, Kabir-2019}, which we follow here.
Using a compartmental framework, each user can be in any of three states: ``\textit{S} - susceptible'' (unaware of the information but potentially receptive), ``\textit{I} - infected'' (aware of the information and willing to spread further), ``\textit{R} - recovered'' (aware but no longer transmitting, e.g. due to information obsolescence).
With only the transitions \textit{S} $\rightarrow$ \textit{I}, \textit{I} $\rightarrow$ \textit{R} allowed, the epidemic process terminates, i.e., the information is no longer being propagated when no infected users remain.
For a given initialisation, the final proportion of recovered users, or equivalently those ever infected by the information, is determined by the basic reproduction number $\mathcal{R}_0$, which represents the average number of new infectors generated by an infected user in a fully susceptible population \cite{Matt-2008}. 

\begin{SCfigure}[][htbp]
    \centering
    \adjincludegraphics[width=0.6\textwidth,trim={{.0\width} {.0\height} {.0\width} {.0\height}},clip]{Figures/fig4.png}
    \caption{\textbf{Simulation of follower count growth using temporal retweet networks (simple contagion framework).} 
    We simulate information cascades from highly aligned users (\stanceA{}, \stanceB{}, uncertain) to swayable users, assuming they are coupled to follower gains for the aligned users.
    To reproduce Figure \ref{fig:obs2}b (solid line), we run simulations (dashed line) on the retweet networks generated from past $n$ months to simulate follower count increase rates in the subsequent 1-month. 
    At each timestamp, we run a grid search for $\mathcal{R}_{t_n}$ over the range $[0, 5]$ with a step size of 0.05, and generate $100$ runs for each value of $\mathcal{R}_{t_n}$ to account for stochasticity. 
    To approximate the likelihood distribution, we show here the top 10\% choices of $\mathcal{R}_{t_n}$ across timestamps that best match the empirical data, with a 1-$\sigma$ error bar.
    \textbf{(a)} $n = 1$
    \textbf{(b)} $n = 3$
    \textbf{(c)} $n = 6$
    \textbf{(d) Parameter choices.} 
    Here $\mathcal{R}_{t_n}$ is related to public interest in the COVID-related topic during the corresponding month, with higher values indicating greater interest. 
    In the early stage, $\mathcal{R}_{t_n}$ often exhibits small peaks during periods of heightened attention (e.g., vaccine rollouts or epidemic surges), as expected.
    As the time approaches when the COVID timeline ceases, $\mathcal{R}_{t_n}$ anomalously exhibits elevated values with increased standard deviation.
    This suggests that the coupling dynamics between retweets and follower growth shift and become noisier during this period, when COVID may no longer be the dominant topic of discussion.
    $\delta$ is the scaling parameter that specifies the proportion of estimated new followers among those infected by the information.
    }
    \label{fig:sim}
\end{SCfigure}

We extend this basic SIR model with variations to accommodate our context under the assumptions stated above, forming the simple contagion framework.
Our framework is grounded in the observations in \S3.2 that COVID-related information is highly time-sensitive, with its impact fluctuating over time \textemdash often closely tied to incidents of varying scales that drive follower gains, while rapidly evolving and becoming obsolete.
As a simplification of this real-world dynamic and inspired by \cite{Cinelli-2020}, we run an SIR process to model a time-dependent information cascade for each sliding one-month window, allowing its $\mathcal{R}_0$ to vary across windows $t_n$ to capture fluctuating public interest around COVID-related topics over time (denoted as $\mathcal{R}_{t_n}$ moving forward).
The differences in follower gains across various content types (\stanceA{}, \stanceB{}, uncertain) are reflected in their temporal retweet networks, where information spreads from highly aligned users to swayable users, which in turn facilitates follower gains.
Specifically, for each sliding 1-month window, we generate three temporal retweet networks \textemdash one for each content type (\stanceA{}, \stanceB{}, uncertain) \textemdash by aggregating retweets from the past $n$ months.
For each of these temporal retweet networks, we initialise by considering the class ``\textit{S}'' as swayable users reachable by highly aligned users, ``\textit{I}'' as highly aligned users that can reach them, and ``\textit{R}'' as none. 
To link follower gains with information flow, we aggregate the follower count of recovered swayable users at the end of the SIR process (when no infected users remain), who are randomly selected from the swayable user pool with a total size determined by the model. 
We consider a proportion of this count, scaled by parameter $\delta$, as the estimated number of new followers.
Note that we keep $\mathcal{R}_{t_n}$ the same across retweet networks of all three content types within a time window, and the scaling parameter $\delta$ constant across all three content types and all time windows. 
This ensures that the relative relationship between the follower gains of each aligned group (e.g., one group growing faster than another) is captured not by tuning parameters, but rather from the temporal retweet networks themselves.
We also choose to calibrate $\mathcal{R}_{t_n}$ separately for each time window rather than inferring it temporally, based on the intuition that the timing and impact of real-world events are inherently unpredictable from past windows (e.g., one cannot know when a vaccine will be officially ruled out before the decision is announced).
Details about the model formulation and parameter optimisation are provided in \S5.2.

Figure \ref{fig:sim}a, b, c show the top 10\% of simulation realisations across parameter choices that best match the empirical data for $n = 1, 3, 6$, respectively. 
We emphasise that our focus below is on whether the relative differences in follower gains between aligned groups can be reproduced at the information flow level.
For all choices of $n$, the follower increase rate for uncertain content is consistently underestimated throughout the timeframe.
This is not surprising, since these users likely have diverse interests beyond COVID (as previously noted), and the followers they attract may largely fall outside those focused on COVID-related content.
In contrast, the trends for \stanceA{} and \stanceB{} content are better captured. 
For $n = 1$, \stanceA{} content is simulated to gain followers more rapidly than \stanceB{} content toward the end of 2020, and the trend is more pronounced than in the empirical data. 
Later, \stanceB{} content is simulated to outperform starting around mid-2021, with this turning point roughly matching the empirical data.
As $n$ increases, the estimated trends become smoother, and the differences in follower count growth between content types become more distinct.
However, this comes at the cost of losing some resolution, particularly where \stanceA{} content temporarily outperforms \stanceB{} content (e.g., late 2020 and early 2022).

We discuss the interpretability of the parameter results below (Figure \ref{fig:sim}d).
The temporal distribution of $\mathcal{R}_{t_n}$ which we relate to the public interest towards COVID topics over time, exhibits small peaks that correspond to local increases in follower gains before the end of 2021.
As expected, higher values of $\mathcal{R}_{t_n}$ often arise during periods of heightened attention, such as vaccine rollouts or surges in epidemic severity.
Yet it is striking to observe several large increases in $\mathcal{R}_{t_n}$ since early 2022, a period when public interest in COVID was expected to decline. 
We also see a larger standard deviation in the $\mathcal{R}_{t_n}$ distribution, with the estimated differences in follower growth between content types becoming less distinguishable, particularly after mid-2022.
This ``anomaly'' likely arises because the coupling dynamics between retweets and follower growth shift and become noisier in the later stage, when COVID may no longer be the dominant topic of discussion.
This is evidenced by the cessation of various COVID timeline documents (top dashed-line region in Figure \ref{fig:sim}), whose discontinuation generally coincides with the onset of the anomaly.
We also observe a similarly anomalous period in the biased convergence framework (see Supplementary Information E).
For the scaling parameter $\delta$, we observe that its value decreases as $n$ increases \textemdash this can be explained by the fact that retweet networks aggregated over a longer time window tend to be denser, allowing information to spread to more users, and thus requiring a smaller $\delta$ to scale the proportion of estimated new followers from those infected.

Finally, we demonstrate in Supplementary Information D that the results in Figure \ref{fig:sim} remain robust, with a distribution similar to the top 5\%.

\section{Discussion}
Amid extraordinary events (e.g., COVID-19) involving large-scale public attention, it is important to maintain online public discourses aligned with reliable, fact-based information.
By drawing on a large-scale temporal dataset of COVID-related retweets, we measure the dynamics of public credibility through the lens of follower growth across news sources with varying levels of scientific quality. 

Over a three-year span from early to late pandemic stages, we observe that users deeply engaged in COVID-related content see rapid visibility gains, marked by follower spikes during major events like vaccine rollouts and case surges.
While credible users (i.e., those retweeting content with higher scientific accuracy) often see a larger boost during these spikes, particularly in the early stages of the pandemic, those doing so with content of more questionable quality tend to maintain faster growth outside of them. 
In contrast, users with uncertain credibility show a more modest increase, with fluctuations less driven by COVID-related events.
These findings suggest a concerning trend: although reliable, evidence-based information may temporarily dominate public attention during moments of high-impact events, lower-quality content can steadily gain traction over time, potentially shaping public perception and behaviour in more subtle, persistent ways. 
This underscores the need for sustained interventions to uphold the visibility of reliable sources, not only during major events but also throughout quieter periods when misinformation is more likely to take hold.

We further support these findings with two scalable modelling frameworks (i.e., simple contagion and biased convergence).
We show that variations in user engagement (as reflected in temporal retweet networks) largely couple to the disparities in visibility (as indicated by follower growth dynamics) between \stanceA{} and \stanceB{} content, though not for uncertain content, as expected.
Our models become less interpretable during the late pandemic stage, likely due to a shift in public interest away from COVID, making much of the follower count growth less coupled to COVID-related retweets.
Our frameworks complement traditional survey data for measuring public credibility in \stanceA{} digital sources versus \stanceB{} ones \textemdash though not replaceable, as surveys more directly capture public opinion \textemdash while our approach gauges credibility through shifts in deliberate attention (i.e., follower count).
By contrast, our measure prevails in integrating public opinion with real-time information traffic and leverages data that is more easily accessible at scale, laying the foundation for an early warning and monitoring system to inform timely interventions and policy design.
Moreover, our modelling frameworks offer an initial step toward linking information flow with follower growth in various temporal social networks (e.g., Bluesky \cite{Smith-2025}), providing insights into public engagement and attention around other polarising topics (e.g., climate change, technological disruption), and serving as a potential benchmark for future research.

\paragraph{Future Work}
Our research opens several avenues for future work.

Here we annotate the scientific quality of retweets based on the web domains they reference.
This is a method that is also commonly used in previous research \cite{Bovet-2019, Flamino-2023, Cinelli-2020}, and in our case potential biases at different levels (e.g., from the Twitter filtering API or manual annotations) have been carefully addressed following \cite{Gallotti-2020} (as mentioned in \S2).
Nevertheless, a key limitation of this method remains: content from the same web domain can vary in quality, and domain-level aggregation may obscure this variation.
For instance, recent literature show that mainstream media (categorised as ``\stanceA{}'' in our case) may still publish misleading content due to novelty or ideological alignment \cite{Tsfati-2020}, and that in some cases misleading content from mainstream outlets is more likely to be shared than content from unreliable sources, contributing to vaccine hesitancy \cite{Allen-2024, Goel-2025}.
One way to tackle this problem in future work is to evaluate the content of the individual articles themselves, which would often require natural language processing (NLP) techniques to automate fact-checking at scale for a dataset of this size.
It is also worth noting that, as Allen et al. \cite{Allen-2024} mention, even at this individual-article level there can be a range of ``gray-area'' content that is ``factually accurate but nonetheless misleading'' (e.g., headlines that are technically true but lack context), which even human fact-checkers may struggle to identify and that can nonetheless reach large audiences.

Our work examines the co-evolution of retweets and follower counts at a coarse scale, both in terms of temporal resolution (1-month time windows) and group dynamics (\stanceA{}, \stanceB{}, and uncertain content).
This scope entails several important limitations and points to finer-grained investigations as directions for future work.
(1) Here we perform network backbone extraction to narrow our focus towards statistically significant retweets that are more likely to lead to follower gains.
With the current dataset, however, causal analysis of this retweet-follower dynamic (e.g., via Granger causality) is not feasible, since follower counts are recorded only at the time of activity (retweet or being retweeted), with an average interval of around 19 days. 
It would be valuable for future work to conduct such analyses using higher-resolution temporal data with regularly monitored follower counts (e.g., every 15 minutes).
(2) Our data timeframe is centred on the COVID-19 pandemic, while also encompassing other major events (e.g., the 2020 US presidential election, the Russia–Ukraine war beginning in 2022).
Although our modelling provides evidence that the ``backboned'' follower count dynamics are coupled to COVID-related retweet dynamics (becoming noisier after early 2022), it is entirely possible that these other events also influenced the follower count dynamics throughout, particularly given the inclusion of mainstream media accounts.
A thorough investigation of the impact of these other events would require, for example, also collecting retweets on these events for the same accounts and conducting parallel causal analyses with follower count dynamics.
This ties back to point 1 made earlier and also lies beyond the scope of the present work, pointing to an interesting direction for future research.
(3) To study group-level dynamics, we aggregate follower counts of highly aligned users by content type (\stanceA{}, \stanceB{}, or uncertain) and track the monthly growth rate of this aggregated count over time, without distinguishing variation at the level of individual users within each group. 
We note, however, that previous literature has shown that individual-level follower growth rates can vary depending on follower base size, which would be worth accounting for in future work.
For instance, Sangiorgio et al. \cite{Sangiorgio-2024} find that accounts with fewer followers tend to grow faster than larger ones until reaching an audience equilibrium, while Johnson et al. \cite{Johnson-2020} observe that medium-sized anti-vaccination pages grow the most compared to pro-vaccination and undecided ones.

We explore two frameworks, ``simple contagion'' and ``biased convergence'', to capture the co-evolving dynamics of retweets and followers. 
In simple contagion, the probability of adoption increases with independent, heterogeneous, and additive exposures, whereas in biased convergence users gradually update their beliefs toward content aligned with their prior bias. 
Results show that simple contagion reproduces more empirical features than the latter. 
We therefore conclude that, in our setting, the dominant dynamics are consistent with a simple-contagion mechanism more capable of explaining heterogeneity, while biased convergence, by construction, emphasises local drift in homophilous environments and reduces cross-cluster propagation, thereby underestimating the scale and variety of the observed diffusion patterns.

\section{Methods}
In this paper, we analyse our dataset from a network perspective, using a specific network type: directed weighted networks, with self-loops and without multi-edges. 
Some preliminary network definitions are listed below. 

\begin{itemize}
    \item A \textit{network} $G = (V, A)$ consists of a set of nodes $V$ and a weighted adjacency matrix $A = (A_{i,j})_{i,j \in V}$. $A_{i,j} = w_{ij} > 0$ if there is an edge from node $i$ to $j$ with weight $w_{ij}$ and $A_{i,j} = 0$ otherwise.
    
    \item A \textit{path} from $i \in V$ to $j \in V$ is defined as a succession of nodes $(n_0, n_1, ..., n_k)$, where $k$ is a non-negative integer, $n_0 = i$, $n_k = j$, and for any $l = 1, ..., k$, $n_l \in V$ are distinct satisfying $A_{n_{l-1},n_{l}} > 0$. As a special case, every node $i \in V$ is considered to have a path $(n_0 = i)$ to itself.
    
    \item Let $i, j \in V$ be nodes and $U \subseteq V$ be a subset of nodes. A node $j$ is said to be \textit{reachable} from $i$ if a path exists from $i$ to $j$. As an extension, $i$ is said to be reachable from $U$ if there exists at least a $u \in U$ such that $i$ is reachable from $u$. $U$ is said to be reachable from $i$ if there exists at least a $u \in U$ such that $u$ is reachable from $i$. 
\end{itemize}

\subsection{Network backbone extraction: disparity filter}
Network backbone extraction encompasses a family of methods aimed at reducing the size of a network by removing nodes or edges from the original structure, while preserving some essential properties. 
A literature review by Yassin et al. \cite{Yassin-2023} classifies network backbone extraction methods into two categories: structural and statistical.
The structural methods seek to preserve specific structural properties (e.g., shortest path \cite{Grady-2012}, spanning tree \cite{Wu-2006}, community structure \cite{Rajeh-2022}).
The statistical methods often retain statistically significant nodes and edges under a null model \cite{Serrano-2009, Marcaccioli-2019, Dianati-2016, Coscia-2017}.

For our interest in this paper, we apply a statistical network backboning method \textemdash disparity filter \cite{Serrano-2009}, which is widely used, suitable to apply in the context of retweet and follower gains dynamics, and computationally affordable on our dataset. 
It aims to preserve edges with a statistically significant high weight at a local scale, compared with a null model where weights are uniformly randomly distributed over its edges. Afterwards, nodes without any linked edges are removed. 
The disparity filter method can be applied to both undirected and directed weighted networks. We describe it below in the context of directed weighted networks, as in our dataset.

\begin{itemize}

\item \textbf{Weight normalisation:} To account for the local fluctuations of weights on the directed edges, a weight normalisation measure is applied first. 
For a directed edge from node $i$ to node $j$ of weight $\omega_{ij}$, its normalised weight relative to all other edges from node $i$ is given by $p_{ij}^{out} = \frac{\omega_{ij}}{\sum_{l}\omega_{il}}$. Similarly, its normalised weight relative to all other edges directed to node $j$ is given by $p_{ij}^{in} = \frac{\omega_{ij}}{\sum_{l}\omega_{lj}}$. 
In this way, each edge, despite having a single weight, is assigned two normalised weights \textemdash one relative to the source node and one to the sink node.

\item \textbf{Null model:} The null model assumes a network with the same connectivity as the empirical one, that is the number of edges from/to each node is the same, but the normalised weights on each edge differ and follow a uniform distribution.
For an edge from node $i$ to node $j$, the probability density function (PDF) of its normalised weight is given by 
$$\rho(x) = (k-1)(1-x)^{k-2}, 0 \leq x \leq 1, k > 1$$
where $k = k_i^{out}$ (out-degree of node $i$, i.e., the number of edges sourced from node $i$), $k_j^{in}$ (in-degree of node $j$, i.e., the number of edges directed towards node $j$) when $\rho(x)$ represents the PDF of $p_{ij}^{out}, p_{ij}^{in}$, respectively. 

\item \textbf{Retaining edges:} We retain an edge from node $i$ to $j$ if its normalised weights, either $p_{ij}^{out}$ or $p_{ij}^{in}$ is considered statistically significant high. That is, at the significance level $\alpha$, all the edges with $\alpha_{ij} < \alpha$  reject the null model and will be retained, where
\begin{gather*}
\alpha_{ij}^{out}= 1 - (k_i^{out}-1)\int_0^{p_{ij}^{out} }(1-x)^{k_i^{out} - 2}dx = (1 - p_{ij}^{out})^{k_i^{out} - 1}\\
\alpha_{ij}^{in}= 1 - (k_j^{in}-1)\int_0^{p_{ij}^{in} }(1-x)^{k_j^{in} - 2}dx = (1 - p_{ij}^{in})^{k_j^{in} - 1}\\
\alpha_{ij} = min\{\alpha_{ij}^{out}, \alpha_{ij}^{in}\}
\end{gather*}
Note that, in the case of a node $i$ of $k_{i}^{out} = 1$ connected to a node $j$ of $k_{j}^{in} > 1$, we keep the edge only if it beats the threshold for node $j$. That is, we assign $\alpha_{ij}^{out} = 1$ and keep the edge when $\alpha_{ij}^{in} < \alpha$. Similarly for the case where a node $i$ of $k_{i}^{out} > 1$ connected to a node $j$ of $k_{j}^{in} = 1$. 
\end{itemize}

The disparity filter method assumes a system with strong disorder, where weights are heterogeneously distributed both globally and locally \textemdash a criterion our dataset meets, as shown in Supplementary Information A.1. 

Due to the large size of our dataset, we implemented the method using parallel programming. 
The code was developed in Python and has been made available for public use. 

\subsection{Simple contagion framework}
The background and rationale for the simple contagion framework have been discussed in \S3.3. 
Below we formulate the model and demonstrate the optimisation of parameter choices.

To reproduce the follower count increase rate $r(t, p)$ of highly aligned users for content types $p = fac, mis, unc$ within a 1-month window $t$ as shown in Figure \ref{fig:obs2}b, we aggregate retweets of content type $p$ from the past $n$ months to generate a retweet network $G(t_n, p)$. 
In $G(t_n, p)$, we denote $V_a(t_n, p)$ as the set of users aligned with content type $p$ who can also reach swayable users. Similarly, we define $V_{sw}(t_n, p)$ as the set of swayable users that $V_{a}(t_n, p)$ can reach. We consider each user $i \in V_a(t_n, p), V_{sw}(t_n, p)$ carries a follower count $f(t_n, i)$, last recorded before the 1-month window $t$.

We initialise the SIR model by considering 
$$S^0(t_n, p) = \frac{|V_{sw}(t_n, p)|}{N(t_n, p)}, \quad I^0(t_n, p) = \frac{|V_a(t_n, p)|}{N(t_n, p)}, \quad R^0(t_n, p) = 0, \quad \text{where} \; N(t_n, p) = |V_{sw}(t_n, p)| + |V_a(t_n, p)|\footnotemark$$
\footnotetext{For any set $A$, we define $|A|$ as its cardinality, i.e., the number of elements in $A$.}

At the end of the epidemic process when no infected users remain, the final proportion of recovered users, or equivalently those ever infected by the information, denoted as $R^\infty(t_n, p)$, is determined by the basic reproduction rate $\mathcal{R}_{t_n}$. It follows the equation below, derived from a standard SIR ODE system by \cite{Matt-2008}.
Note that we keep $\mathcal{R}_{t_n}$ the same for all content types $p$, but allow it to vary across different time windows. 
$$1 - R^\infty(t_n, p) - S^0(t_n, p)e^{-R^\infty(t_n, p)\mathcal{R}_{t_n}} = 0$$
Given $\mathcal{R}_{t_n}$ and $S^0(t_n, p)$, an explicit solution for $R^\infty(t_n, p)$ cannot be found, as it is a transcendental equation. 
We obtain an implicit solution using the Newton-Raphson algorithm \textemdash a numerical method mentioned by \cite{Matt-2008}. 
In the following, we denote $R^\infty(t_n, p)$ as $R^\infty(t_n, p \,|\, \mathcal{R}_{t_n})$ to indicate its dependence on the choice of $\mathcal{R}_{t_n}$. 

By excluding aligned users from the set of the recovered users, we identify the set of recovered users who are swayable \textemdash specifically, the number of users in this subset:
$$N_{sw-rec}(t_n, p \,|\, \mathcal{R}_{t_n}) = N(t_n, p)(R^\infty(t_n, p \,|\, \mathcal{R}_{t_n}) - I^0(t_n, p))$$

We now proceed to associate the SIR model with the follower gains. 
From the pool of $V_{sw}(t_n, p)$, we select $N_{sw-rec}(t_n, p \,|\, \mathcal{R}_{t_n})$ number of users following a uniform distribution to form the set $V_{sw-rec}(t_n, p\,|\, \mathcal{R}_{t_n})$. Their aggregated followers are then considered as those attracted to content type $p$ and intending to become new followers of the corresponding highly aligned users. We scale a fraction of them by $\delta$ to estimate the number of users who actually become new followers of the highly aligned users. 
Note that we keep the scaling parameter $\delta$ constant across all content types $p$ and time windows $t_n$. Then the estimated follower count increase rate within the 1-month window is calculated as 
$$\hat{r}(t_n, p\,|\, \delta, \mathcal{R}_{t_n}) = \frac{\sum\limits_{i \in V_{sw-rec}(t_n, p\,|\, \mathcal{R}_{t_n})} \delta f(t_n, i)}{\sum\limits_{i \in V_a(t_n, p)} f(t_n, i)}$$

To reproduce Figure \ref{fig:obs2}b with a given $n$, we minimise the sum of squared differences between $\hat{r}(t_n, p)$ and $r(t_n, p)$ over all time windows $t_n$ and content types $p$:
$$
Q(\delta,\mathcal{R}_{t_n} \text{ for all } t_n) 
= \sum_{t_n} q(t_n\,|\,\delta, \mathcal{R}_{t_n})
= \sum_{t_n}\sum_{p}(\hat{r}(t_n, p\,|\, \delta, \mathcal{R}_{t_n}) - r(t, p))^2
$$
Here the parameters to optimise include $\mathcal{R}_{t_n}$ over all time windows $t_n$, as well as the scaling parameter $\delta$. 
Note that for a given $\delta$, minimising $Q$ is equivalent to minimising $q$ within each time window $t_n$, as $\mathcal{R}_{t_n}$ for each $t_n$ is chosen independently. 

For each time window $t_n$, we run a grid search for $\mathcal{R}_{t_n}$ over the range $[0, 5]$ with a step size of $0.05$. 
We see in Figure 2.2 of \cite{Matt-2008} that this range covers the recovered user fraction from $0$ to nearly $1$ when $S^0(t_n, p) = 1$. 
To account for stochasticity in selecting recovered swayable users, we run $100$ simulations for each value of $\mathcal{R}_{t_n}$. 
Borrowing the idea of widely used Approximate Bayesian Computation \cite{Beaumont-2002} to approximate the likelihood function of $\mathcal{R}_{t_n}$, we set a tolerance percentage for $q$ (e.g., smallest 10\% in Figure \ref{fig:sim}) and accept values of $\mathcal{R}_{t_n}$ that fall below the threshold. 
The scaling parameter $\delta$ is then obtained using the Nelder–Mead numerical method by minimising the sum of $q$ over all accepted choices of $\mathcal{R}_{t_n}$ and over all time windows $t_n$. 

\vspace{5mm}
\paragraph{Data \& code availability}
The dataset used in this study cannot be made publicly available due to privacy regulations. 
However, we provide the tweet IDs of the collected data, allowing anyone to retrieve the tweets directly via Twitter's API.
Alternatively, the full dataset can be obtained from the corresponding authors upon reasonable request.
Both the tweet IDs and the programming code used in this study are stored in \url{https://github.com/YuetingH/COVID_Retweets}.

\paragraph{Acknowledgements} 
Y.H. acknowledges support from EPSRC grant no. EP/S022244/1 through the MathSys CDT, University of Warwick.
P.T. acknowledges support from the Leverhulme Trust for the Research Grant RPG-2023-050 titled ``Promoting Social Good Using Social Networks''. 
G.A. acknowledges support from FAIR: Future Artificial Intelligence Research - PNRR MUR.

\paragraph{Competing interests}
The authors declare no competing interests.

\bibliographystyle{unsrt}
\bibliography{bibliography.bib}

\newpage
\appendix
\section*{Supplementary Information} 
\subsection*{A. Disparity filter}

\subsubsection*{A.1. Assumption of heterogeneity}
The disparity filter method, proposed by Serrano et al. \cite{Serrano-2009}, works in systems with strong disorder, where network edge weights are distributed heterogeneously at both global and local levels. Here we show our dataset satisfies this assumption, which we examine in the same way as Serrano et al. (see Figure 5 in their paper). The notations below remain consistent with those used in our main paper.

\begin{figure}[ht]
    \centering
    \adjincludegraphics[width=.9\textwidth,trim={{.00\width} {.00\height} {.04\width} {.05\height}},clip]{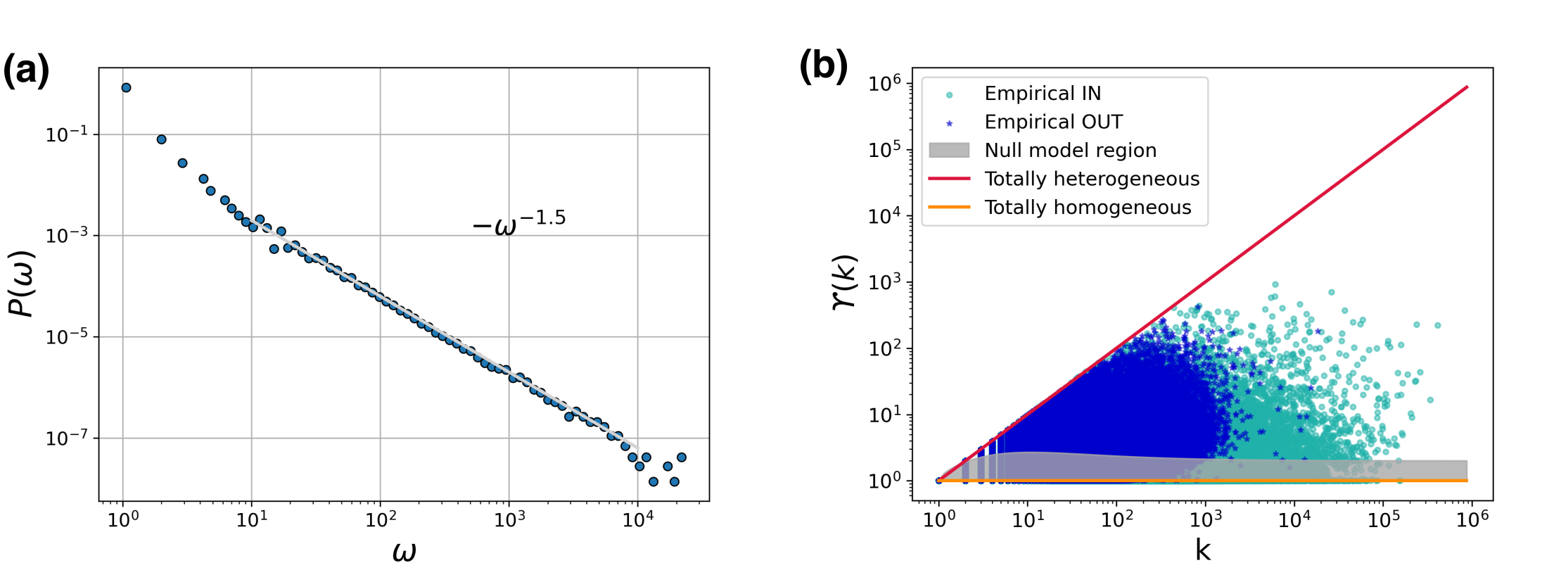}
    \caption{\textbf{Heterogeneity at the (a) global (b) local level. }} 
    \label{fig:SM1}
\end{figure}

At the global level, the heterogeneity is evaluated through the distribution of edge weights. 
Figure \ref{fig:SM1}a reveals that the distribution of weights in our dataset is heavy-tailed \textemdash a power-law fit of the form $\omega^ {-\beta}$ yields an exponent $\beta = 1.5$. 

At the local level, the heterogeneity of each node $i$ with in-degree $k_i^{in}$ and out-degree $k_i^{out}$ is measured as below:
\begin{gather*}
\Upsilon_i(k_i^{in}) = k_i^{in}\sum_j(p_{ij}^{in})^2 \in [1, k_i^{in}], \text{where } k_i^{in} \geq 1 \\
\Upsilon_i(k_i^{out}) = k_i^{out}\sum_j(p_{ij}^{out})^2 \in [1, k_i^{out}], \text{where } k_i^{out} \geq 1 
\end{gather*}
Under perfect homogeneity, when all edges share the same weight, $\Upsilon_i(k_i^{in}) = 1$ ($\Upsilon_i(k_i^{out}) =1$). 
Under perfect heterogeneity, when a single edge connected to the node carries all the weight of its attached edges, $\Upsilon_i(k_i^{in}) = k_i^{in}$ ($\Upsilon_i(k_i^{out}) = k_i^{out}$).
In the case of the null model (as described in the Methods), the average and the variance of the heterogeneity are found to be:
\begin{gather*}
\mu(\Upsilon_{null}(k)) = \frac{2k}{k+1} \\
\sigma^2(\Upsilon_{null}(k)) = k^2(\frac{20+4k}{(k+1)(k+2)(k+3)} - \frac{4}{(k+1)^2}) \\
\text{where } k\geq1 \text{, } k = k_i^{in} \text{ or } k_i^{out}
\end{gather*}
The local heterogeneity will be recognised only if the observed values lie outside this area: $\Upsilon_{ob}(k) > \mu(\Upsilon_{null}(k)) + a\cdot\sigma(\Upsilon_{null}(k))$.
We demonstrate in Figure \ref{fig:SM1}b that our dataset exhibits strong local heterogeneity, where we choose $a =2$ (the same as Serrano et al.).

\subsubsection*{A.2. Choosing significance level $\bm{\alpha}$}
\begin{figure}[ht]
    \centering
    \adjincludegraphics[width=.4\textwidth,trim={{.00\width} {.05\height} {.00\width} {.00\height}},clip]{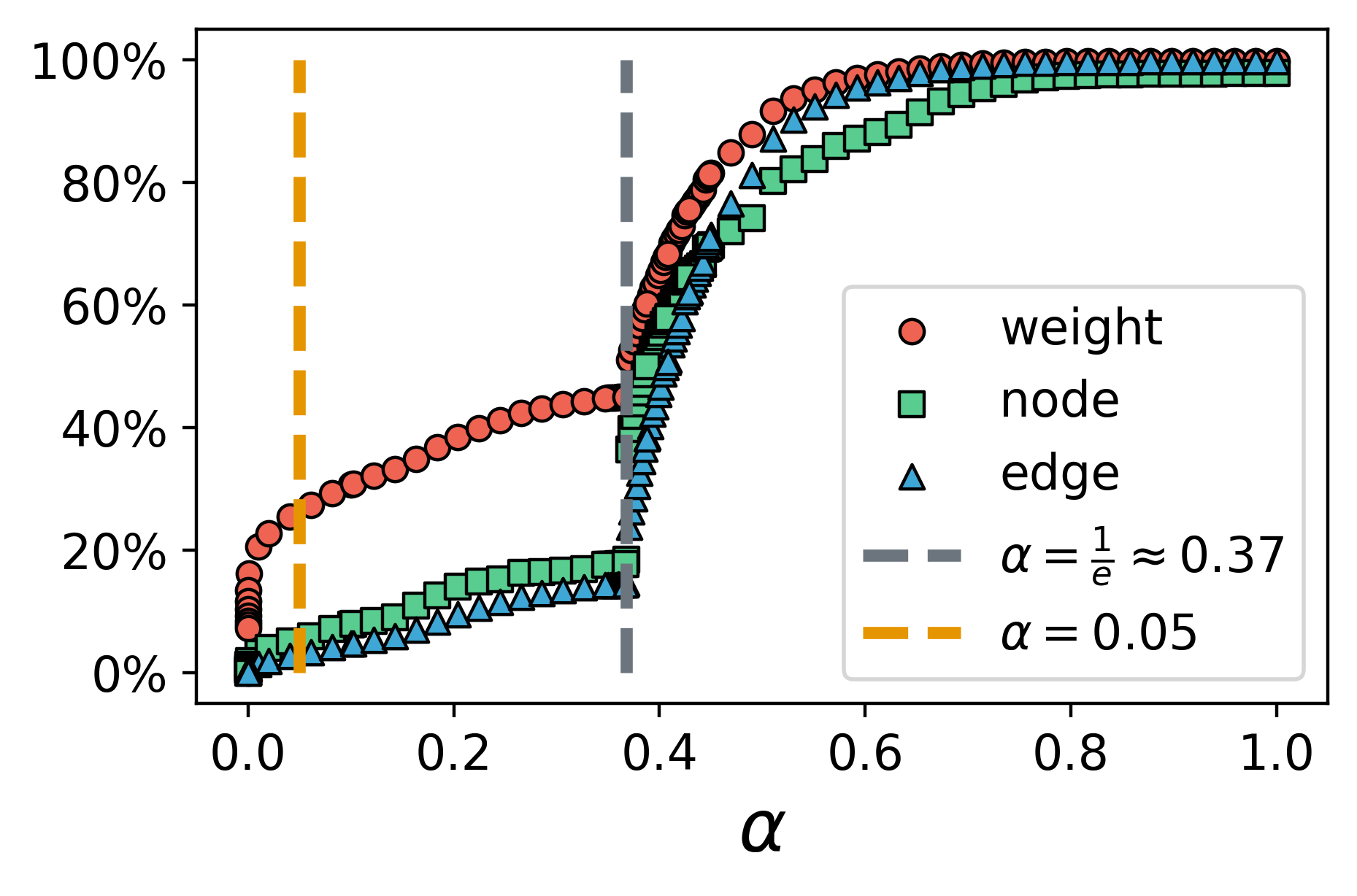}
    \caption{\textbf{Sizes of disparity backbones for different significance levels $\bm{\alpha}$.} For each value of $\alpha$, we calculate the fraction of weight, nodes, and edges kept in the backbones compared to the original network. } 
    \label{fig:SM2}
\end{figure}

For different significance levels $\alpha$, we show in Figure \ref{fig:SM2} the sizes of the network backbones retained by the disparity filter. 
Note that there exists a shape cutoff around $\alpha = \frac{1}{e} \approx 0.37$, which indicates that a large number of edges connects many nodes with nearly equal weights.
In other words, many users give or receive a large number of retweets evenly from others \textemdash a scenario potentially suggesting that one user's content is not of particular interest to another, making it less likely to attract followers. This is the type of case we aim to filter out.
Mathematically, when $k_i^{out} \rightarrow \infty$ ($k_i^{in} \rightarrow \infty$), $p_{ij}^{out} \rightarrow \frac{1}{k_i^{out}}$ ($p_{ij}^{in} \rightarrow \frac{1}{k_i^{in}}$), we obtain $\alpha_{ij}^{out} = (1 - p_{ij}^{out})^{k_i^{out} - 1} \rightarrow \frac{1}{e}$ ($\alpha_{ij}^{in} \rightarrow \frac{1}{e}$). 

\begin{figure}[!ht]
    \centering
    \adjincludegraphics[width=.75\textwidth,trim={{.00\width} {.00\height} {.00\width} {.00\height}},clip]{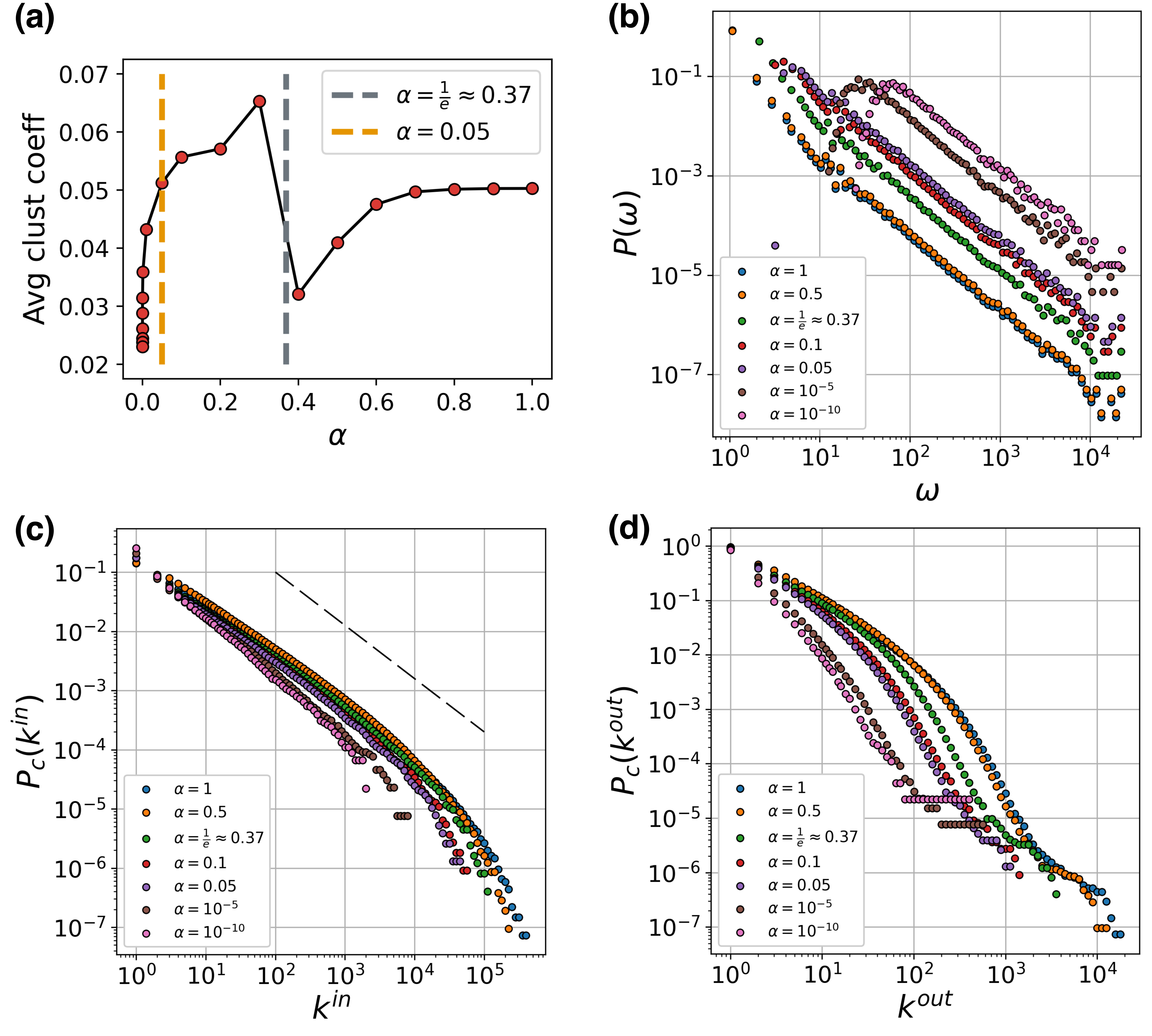}
    \caption{\textbf{Topological properties of disparity backbones for different significance levels $\bm{\alpha}$. (a) Average clustering coefficient. (b) Edge weight distribution.} For $\alpha \geq 0.05$, the filtered network maintains weight distributions that resemble those of the original network. However, for $\alpha = 10^{-5}, 10^{-10}$, this congruence fails for the segment involving small weights. This suggests that filters with extremely small $\alpha$ may excessively remove a large number of edges with small weights, which could still be statistically significant at the local scale.
    \textbf{(c), (d) Complementary cumulative degree distribution. } The findings remain similar to (b). Interestingly, the in-degree distribution follows a power-law pattern, with the power-law exponent $\beta - 1 = 0.9$, whereas the out-degree exhibits a concave log-log distribution. This is possibly because sources with high out-degree, while serving as statistically significant spreaders of COVID-related content, often distribute various types of news content, which we exclude for our interest in this paper.} 
    \label{fig:SM3}
\end{figure}

\begin{figure}[ht]
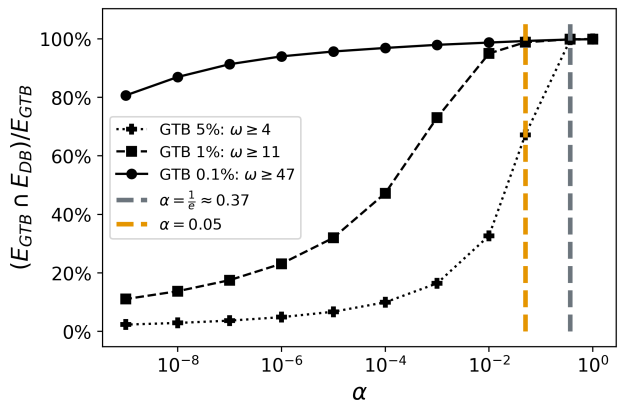

    \centering
    \adjincludegraphics[width=.5\textwidth,trim={{.00\width} {.03\height} {.00\width} {.01\height}},clip]{Figures/SM_Fig4.png}
    \caption{\textbf{Fraction of edges in different global threshold backbones (GTB) included in the disparity backbone (DB) as a function of the significance level.}} 
    \label{fig:SM4}
\end{figure}

Figure \ref{fig:SM3}a demonstrates that $\alpha = 0.05$ (i.e., the choice made in our main paper) keeps the filtered network's average clustering coefficient \cite{Fagiolo-2007} close to the original network, while minimising the network's size. 
Note that here we consider the clustering coefficient that disregards edge weights, as the size of our dataset makes the weighted version computationally infeasible.
Our choice of $\alpha = 0.05$ is further supported when examining other topological properties (see Figure \ref{fig:SM3}b, c, d). 

We also show in Figure \ref{fig:SM4} that this choice effectively preserves most edges with the highest 1\% and 0.1\% weights. Here, the global threshold backbone (GTB) refers to a simple network backbone extraction method that eliminates all edges with weights below a pre-defined threshold.

\subsubsection*{A.3. Retention of bot, verified users and retweets across categories (ungeneralised) at $\bm{\alpha = 0.05}$}
\begin{figure}[ht]
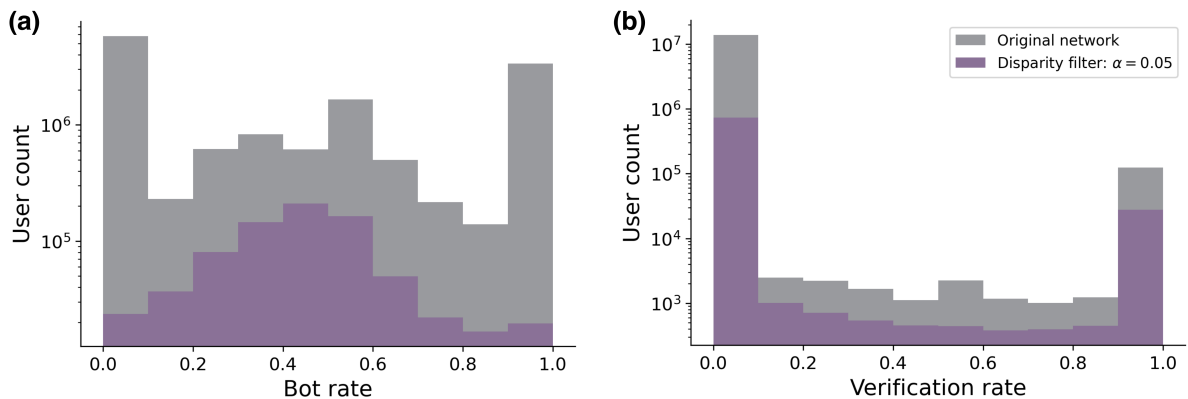

    \centering
    \adjincludegraphics[width=.95\textwidth,trim={{.00\width} {.03\height} {.00\width} {.00\height}},clip]{Figures/SM_Fig5.png}
    \caption{\textbf{Distribution of user (a) bot rate and (b) verification rate at $\bm{\alpha = 0.05}$. (a) } The bot rate distribution for the filtered network at  $\alpha = 0.05$ is left-skewed, but not in the original network. Additionally, the original network's distribution has high bars on both ends because more than 70\% of these bars represent users who either retweeted or were retweeted only once throughout the entire data timeframe, resulting in bot rates of either 1 or 0. The filtered network, on the other hand, removes all such nodes. \textbf{(b)} 99.8\% of users maintained a consistent verification status throughout the entire data timeframe, resulting in two high bars on both ends of the distribution. Of the consistently verified users in the original network (the rightmost bar), around 21.8\% were retained by the filtered network at $\alpha = 0.05$. In contrast, for consistently unverified users (the leftmost bar), only approximately 5.3\% were retained by the filtered network.} 
    \label{fig:SM5}
\end{figure}

We investigate the extent to which bot and verified users are filtered out by the disparity filter at $\alpha = 0.05$. 
Note that bot detection and verification status checks are performed each time a user's activity (i.e., retweeting or being retweeted) is recorded in our dataset. 
As these statuses may change over time, we instead measure the proportion of times a user was detected as a bot across all detections, and the proportion of times their recorded status was verified across all records. 
We refer to these as the ``bot rate'' and ``verification rate'', respectively. Figure \ref{fig:SM5}a indicates that, in general, more bot users were filtered compared to non-bot users. 
This observation is supported by the analysis of the verification status of the Twitter users in Figure \ref{fig:SM5}b, where more verified users were retained.

\begin{figure}[ht]
    \centering
    \adjincludegraphics[width=.85\textwidth,trim={{.00\width} {.02\height} {.00\width} {.00\height}},clip]{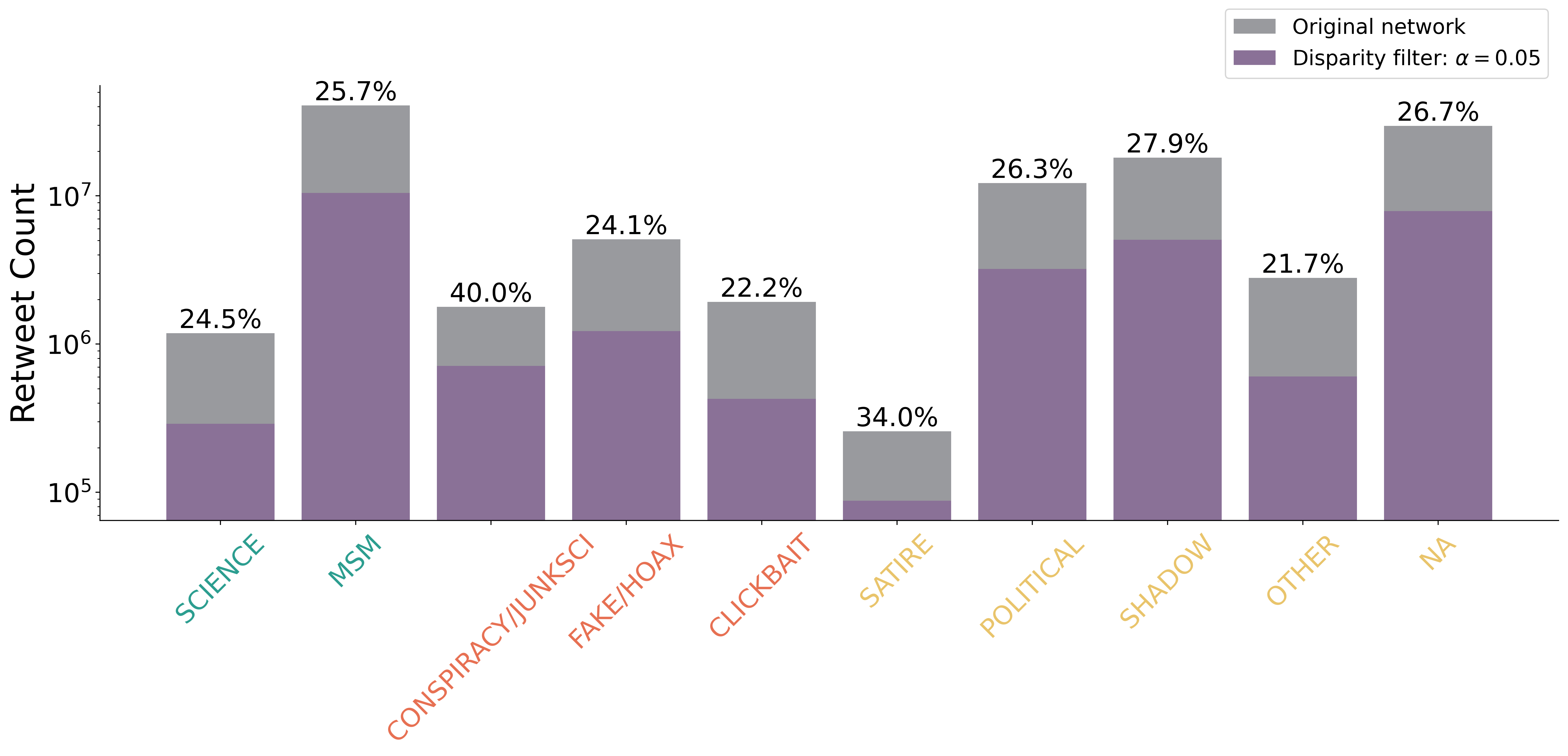}
    \caption{\textbf{Distribution of retweet categories (ungeneralised) at $\bm{\alpha = 0.05}$.} The category text is colour-coded to represent generalised classifications: green for factual, red for misleading, and yellow for uncertain, consistent with our main paper.}  
    \label{fig:SM6}
\end{figure}

We show in Figure \ref{fig:SM6} the distribution of retweet categories (ungeneralised) retained, again at $\alpha = 0.05$. 
Interestingly, we find that retweets categorised as ``CONSPIRACY/JUNKSCI'' (misleading) and ``SATIRE'' (uncertain) exhibit the highest retention percentage, while those categorised as ``CLICKBAIT'' (misleading) and ``OTHER'' (uncertain) show the lowest. 

\subsection*{B. Distribution of user follower counts at the individual level}

\begin{figure}[ht]
    \centering
    \adjincludegraphics[width=.6\textwidth,trim={{.00\width} {.03\height} {.00\width} {.00\height}},clip]{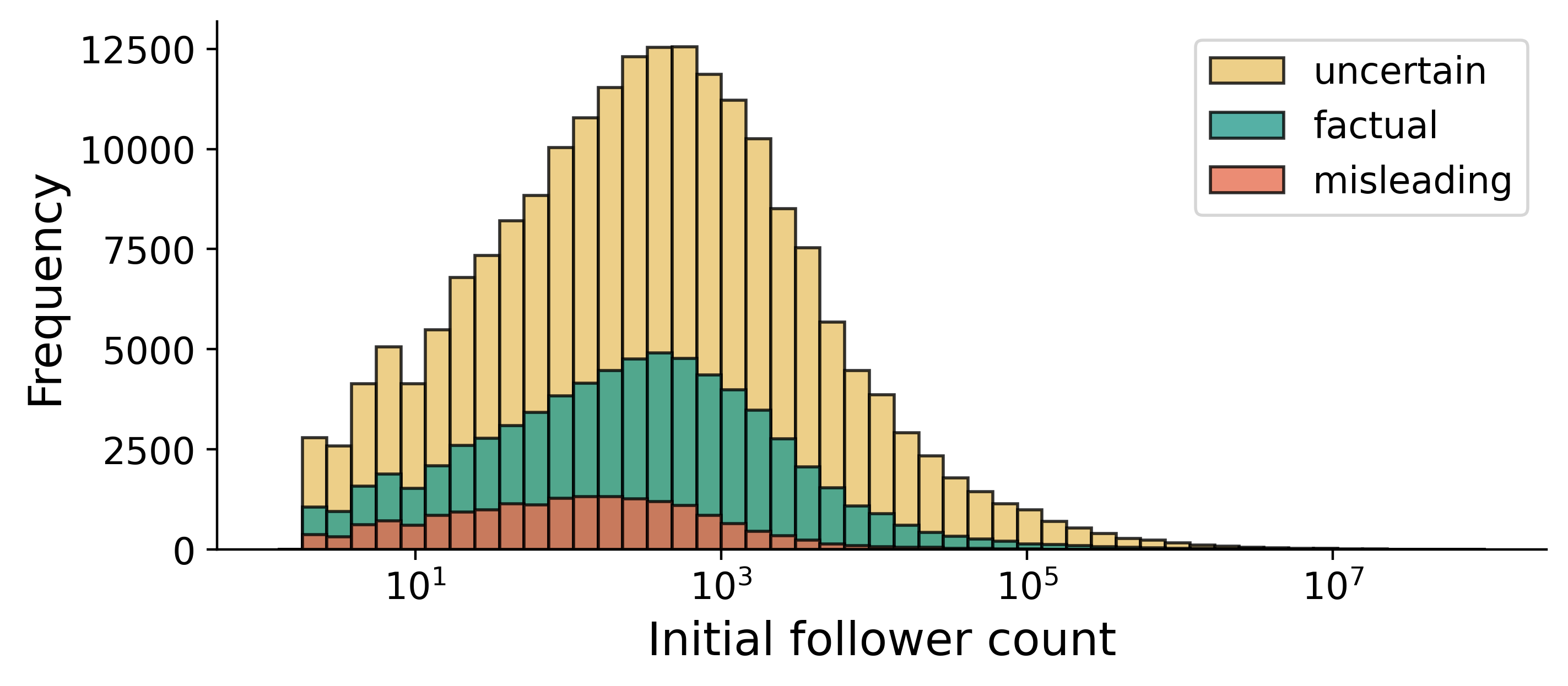}
    \caption{\textbf{Initial follower count distribution of highly aligned users.}} 
    \label{fig:SM7}
\end{figure}

Figure \ref{fig:SM7} depicts the distribution of highly aligned users' follower counts at the start of the dataset (17 March 2020), i.e., their first recorded follower count.
At this initial point, the total follower counts of users aligned with 
factual, misleading, and uncertain content are approximately 929.9 million, 43.6 million, and 3.0 billion, respectively.

\subsection*{C. Creators \& consumers of fake news following Castioni et al. \cite{Castioni-2022}}

Below we show that our dataset exhibits a pattern consistent with that observed by Castioni et al. \cite{Castioni-2022} for an earlier stage of the data: most fake news appears to be created by a minority of users but consumed by a majority.
Note that what is referred to as ``fake news'' in their study corresponds to ``\stanceB{} content'' in our work.
For ease of comparison, we refer to it as ``fake news'' in this section.

\begin{figure}[htbp]
    \centering
    \adjincludegraphics[width=.6\textwidth,trim={{.00\width} {.03\height} {.00\width} {.00\height}},clip]{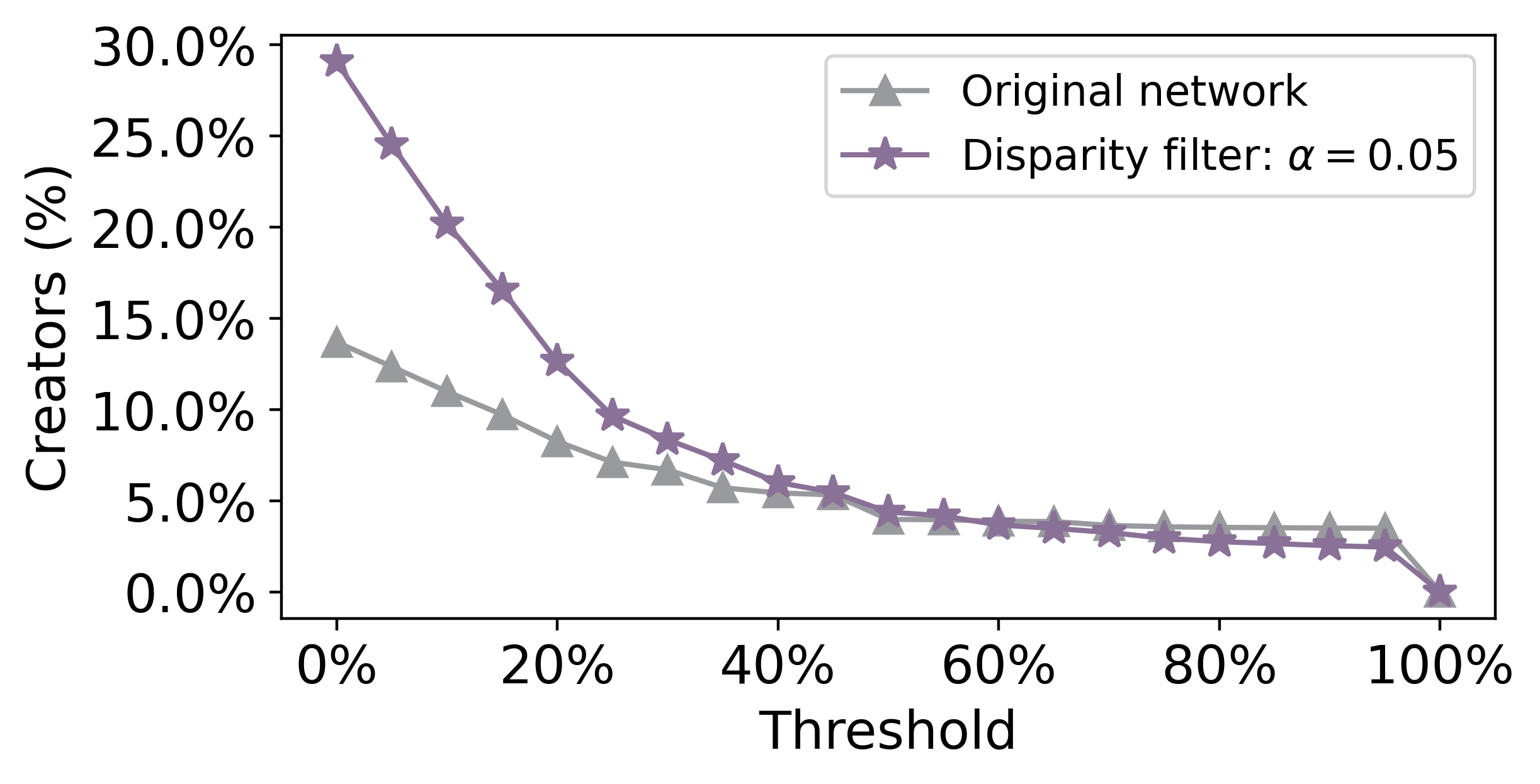}
    \caption{\textbf{Creators (and consumers) of fake news as defined in Castioni et al. \cite{Castioni-2022}.} Users not identified as creators are classified as consumers.} 
    \label{fig:SM8}
\end{figure}

We apply the same criterion as used in \cite{Castioni-2022}, based on the fraction of fake news produced: ``if among all the tweets that an account has produced 20\% or more is fake, then that user is considered to be a creator. On the other hand, if this fraction is between 0\% and 20\%, the account is considered to be a consumer.''
The results are consistent with \cite{Castioni-2022} for both the original and filtered networks ($\alpha = 0.05$).
In our dataset, 8\% of users are classified as creators (92\% consumers) in the original network and 13\% (87\% consumers) in the filtered network, compared with 14\% creators (86\% consumers) in their dataset.
Similar to \cite{Castioni-2022}, this result is robust across different choices of thresholds (see Figure \ref{fig:SM8}). 

\subsection*{D. Simple contagion framework}
\begin{SCfigure}[][htbp!]
    \centering
    \adjincludegraphics[width=0.67\textwidth,trim={{.0\width} {.0\height} {.0\width} {.0\height}},clip]{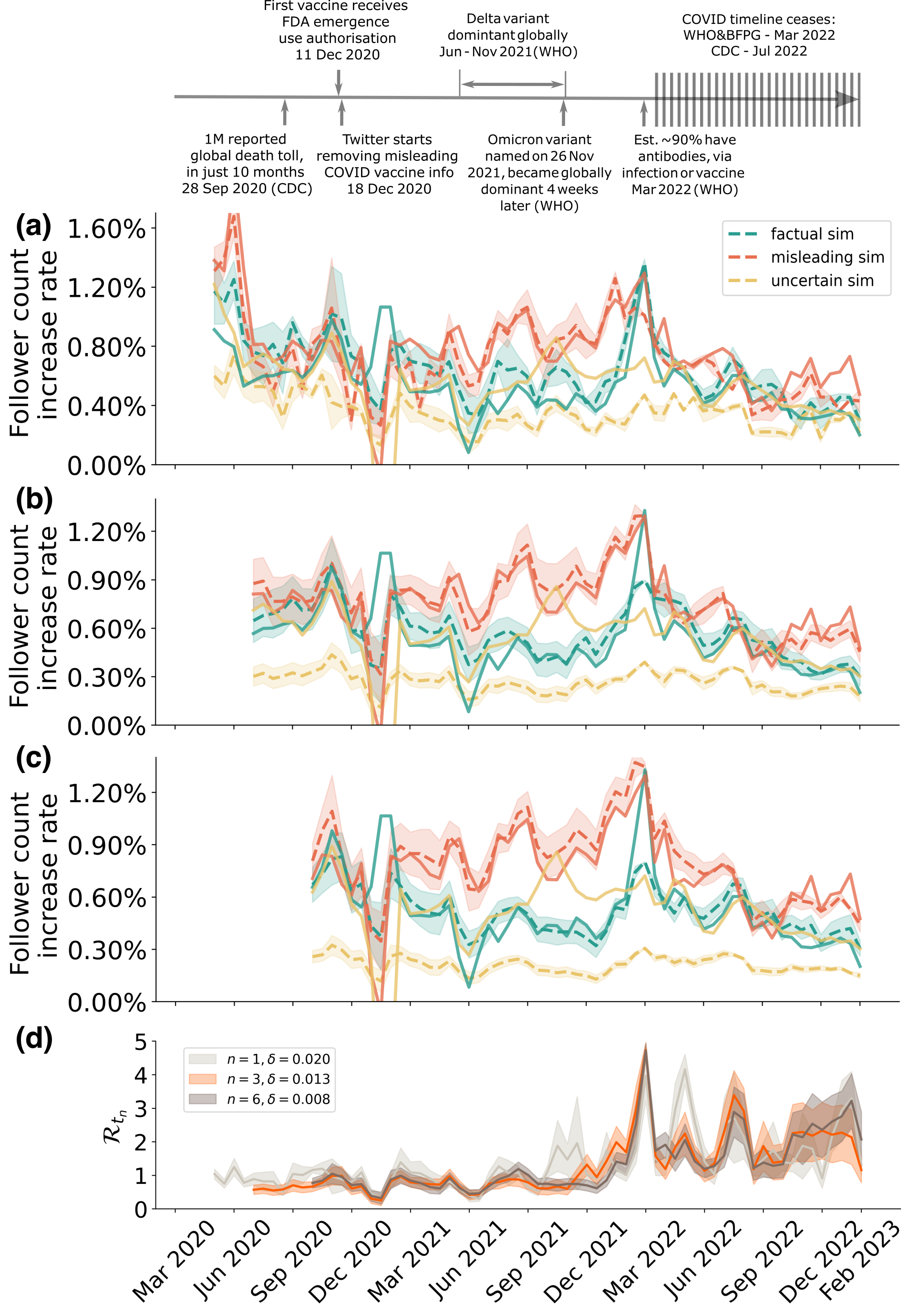}
    \caption{\textbf{Simulation of follower count growth using temporal retweet networks (simple contagion framework - top 5\%).} 
    \textbf{(a)} $n = 1$
    \textbf{(b)} $n = 3$
    \textbf{(c)} $n = 6$
    \textbf{(d) Parameter choices.} 
    }
    \label{fig:SM9}
\end{SCfigure}

\begin{SCfigure}[][htbp!]
    \centering
    \adjincludegraphics[width=0.67\textwidth,trim={{.0\width} {.0\height} {.0\width} {.0\height}},clip]{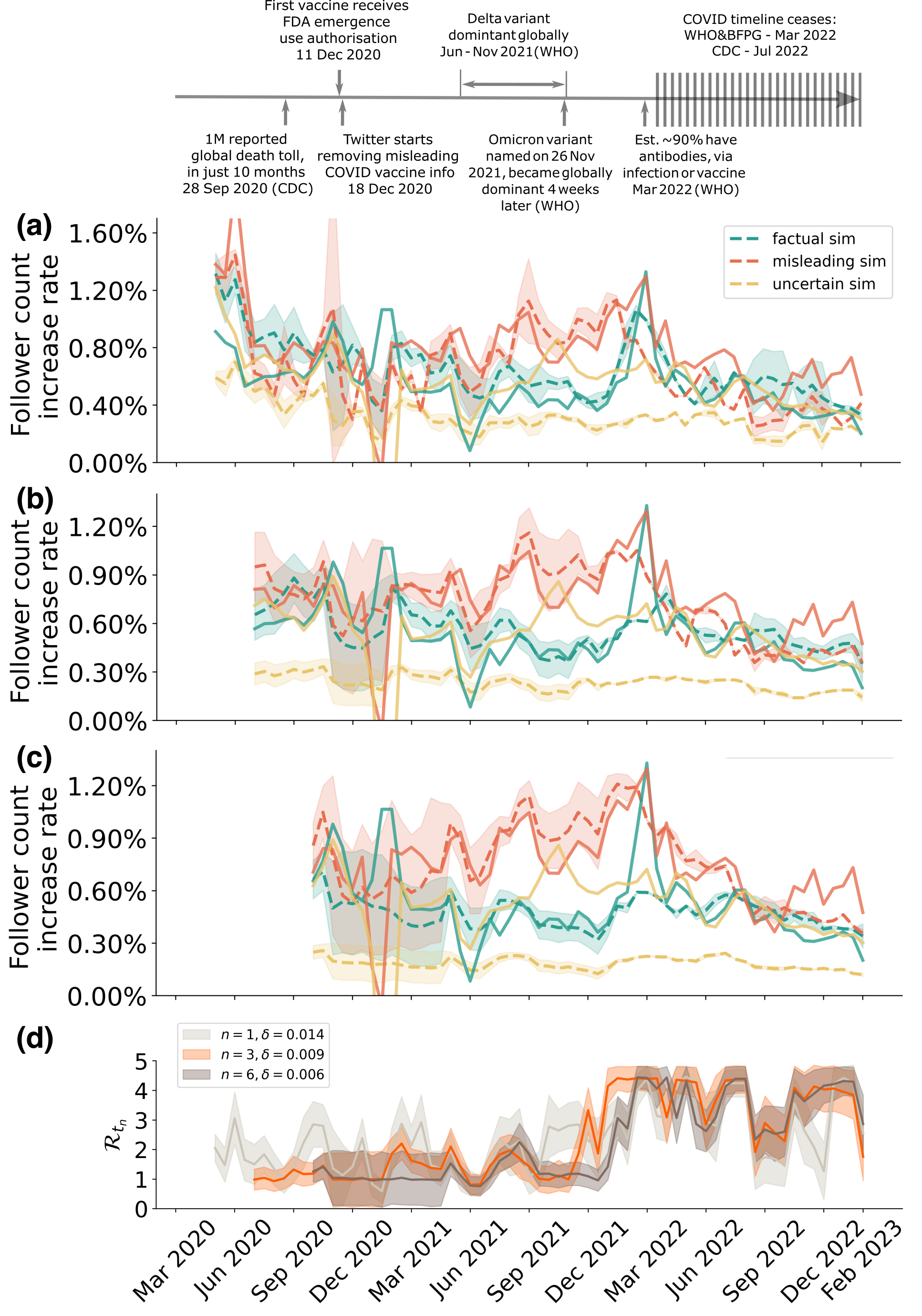}
    \caption{\textbf{Simulation of follower count growth using temporal retweet networks (simple contagion framework - top 20\%).} 
    \textbf{(a)} $n = 1$
    \textbf{(b)} $n = 3$
    \textbf{(c)} $n = 6$
    \textbf{(d) Parameter choices.} 
    }
    \label{fig:SM10}
\end{SCfigure}

In contrast to Figure 4 in the main paper, which presents the top 10\% of simulation realisations across parameter choices that best match the empirical data of follower growth, here we show the results for the top 5\% and 20\% in Figures \ref{fig:SM9} and \ref{fig:SM10}, respectively.

We observe that Figure \ref{fig:SM9} remains very similar to Figure 4, with a slightly closer match to the empirical data. 
As expected, both the $\mathcal{R}_{t_n}$ and the follower count growth rates exhibit slightly smaller standard deviations than Figure 4.
Figure \ref{fig:SM10} still broadly aligns with the empirical data but shows some divergence, particularly toward the end of 2020, possibly due to Twitter's interventions in removing COVID-related misinformation \textemdash a human-driven action that can hardly be captured from historical retweet patterns.

\subsection*{E. Biased convergence framework}

To simulate the coupling dynamics between retweets and follower growth, the main paper uses a simple contagion framework inspired by the epidemic SIR model, whereas here we develop a biased convergence framework based on the classic Friedkin–Johnsen (FJ) opinion dynamics model \cite{Friedkin-1990}.

We first introduce the FJ model, one of the few opinion dynamics models that has been ``experimentally validated for small and medium-size groups'' \cite{Friedkin-1990, Proskurnikov-2017}.
In this model, each agent holds a continuous opinion (typically from 0 to 1, with higher values indicating more extreme opinions) that evolves over discrete time steps, by balancing their own opinion with those of their neighbours.
Each agent carries two parameters: an initial prejudice (its starting opinion before any interaction) and a susceptibility (ranging from 0 to 1, indicating its sensitivity to neighbours' opinions, with higher values indicating higher sensitivity).
Under certain criteria, the system can reach a convergence state.
One often considers the case where ``stubborn'' agents exist, that is, agents that are not sensitive to the opinions of their neighbours at all, with susceptibility set to 0.

We develop the biased convergence framework using the same sliding-window technique as in the simple contagion framework, replacing the epidemic model with the FJ model, and no window-specific calibration parameters (i.e., $\mathcal{R}_{t_n}$) are involved. 
Under this framework, we run the FJ model under two scenarios: with memory (scenario 2) across timestamps (initialising each timestamp with the previous timestamp's converged opinions) and memoryless (scenario 1). 
Specifically, to simulate follower gains for each sliding 1-month window, we run an FJ model on three temporal retweet networks - one for each content type (factual, misleading, uncertain), by aggregating retweets of the corresponding type from the past $n$ months. 
Inspired by Out et al. \cite{Out-2024}, who study the impact of biased media sources using the FJ model, we adopt the following setup in each FJ model run to couple retweet and follower dynamics.
Each highly aligned user is fully stubborn (i.e., susceptibility 0) and holds an extreme opinion at all times (i.e., prejudice 1).
Each swayable user, influenced by highly aligned users (with uniformly distributed susceptibility, constant over all time windows), has two initial-prejudice scenarios: (1) memoryless - prejudice set to 0, and (2) with memory - prejudice equal to their convergent opinion from their last active window (if any, otherwise 0). 
From each swayable user, follower gains in a given time window are proportional to the increase (if any) in their convergent opinion over their initial prejudice, multiplied by their initial follower count in that window.
The simulation involves only a single scaling parameter $\delta$, whose design follows that of the simple contagion framework and remains constant across all time windows and content types.
Further details regarding the model formulation are provided in Supplementary Information E.1.

Figure \ref{fig:SM11} shows 20 realisations for each of the two scenarios of the biased convergence framework with $n = 1$, while simulations for larger window sizes ($n = 3, 6$) are computationally too expensive for our dataset.
Overall the simulations in both scenarios show poor agreement with the empirical data, especially noting that the simulation plots (Figure \ref{fig:SM11}a, b) use a log-scale y-axis, whereas the empirical plot (Figure \ref{fig:SM11}c) does not.
These simulations capture (though overestimating) the leading trend for \stanceB{} content but fail to reproduce the temporary overtakes by \stanceA{} content, except around March 2022.
Scenario 1 slightly better reproduces the local curved trend shapes of \stanceA{} and \stanceB{}, especially in 2021, but shows a long-term trend opposite to the empirical pattern, with a smaller initial increase followed by a larger increase toward the end.
In contrast, scenario 2 better aligns with the empirical long-term trend, showing larger follower growth at the beginning of 2020.
This is expected, as it carries memory of converged opinions across time windows, and users initially have no prior memory, resulting in a larger gap from their initial prejudice and thus higher follower growth.

These results from the biased convergence framework provide a comparison that supports the design and findings of the simple contagion framework in our main paper.
(1) The simple contagion framework yields better results than the biased convergence framework, as it places the follower growth rates of all three content types on a similar scale aligned with the empirical data, especially for \stanceA{} and \stanceB{} content.
(2) We also observe from the biased convergence framework that the local trends in follower growth (in terms of shape rather than scale) deviate somewhat from the empirical data, suggesting that retweet dynamics alone may not be sufficient to explain follower growth fluctuations over time.
This motivates the inclusion of $\mathcal{R}_{t_n}$, which we relate to fluctuating public interest in COVID-related topics over time, as a calibration parameter in the simple contagion framework.
(3) Although the biased convergence framework does not underestimate the follower growth of uncertain content as the simple contagion framework does, it remains consistent with it in showing much less fluctuation for uncertain content than for \stanceA{} and \stanceB{} content.
This supports our interpretation that these users likely have more diverse interests beyond COVID and attract followers less focused on COVID-related content.
(4) As in the simple contagion framework, the biased convergence framework exhibits anomalously large fluctuations and elevated standard deviations in follower growth starting around the end of the COVID timeline, whereas \stanceA{} and \stanceB{} content exhibit more synchronised fluctuations beforehand.
This supports our interpretation that the coupling between retweets and follower growth shifts and becomes noisier in the later stage, as COVID ceases to be the dominant topic of discussion.

\begin{SCfigure}[][htbp!]
    \centering
    \adjincludegraphics[width=0.67\textwidth,trim={{.03\width} {.05\height} {.05\width} {.02\height}},clip]{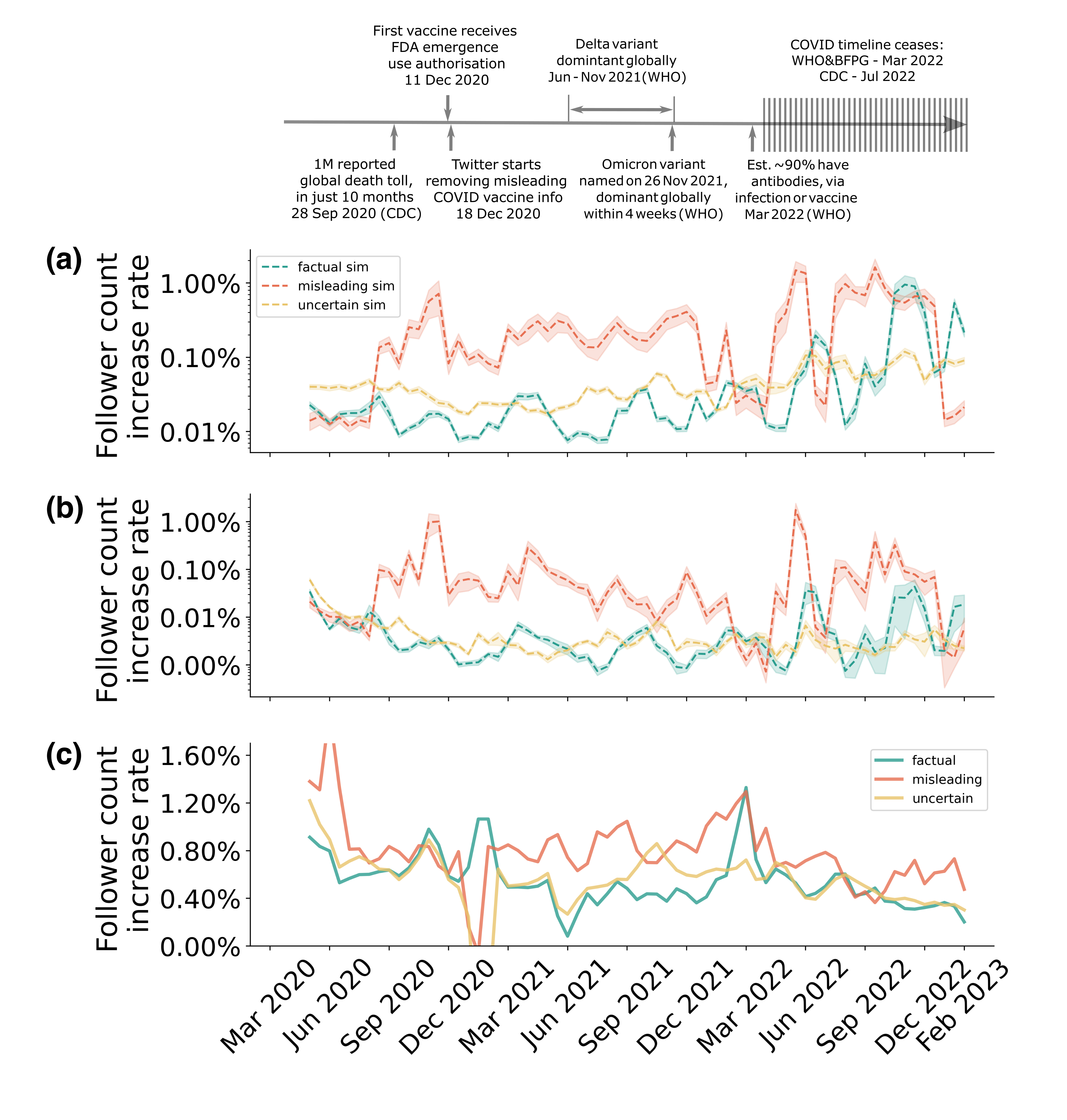}
    \caption{\textbf{Simulation of follower count growth using temporal retweet networks (biased convergence framework - $\mathbf{n = 1}$).} 
    We run 20 realisations for each scenario to account for stochasticity, and show the averages as dashed lines with 1$\sigma$ error bars. $\delta$ is the scaling parameter, whose design follows that of the simple contagion framework.
    \textbf{(a) Scenario 1: memoryless.} $\delta = 1.1 \times 10^{-4}$. 
    \textbf{(b) Scenario 2: with memory.} $\delta = 1.7 \times 10^{-4}$.
    \textbf{(c) Empirical data.} Same as Figure 3(b) in the main paper, included here for comparison with the simulation results.
    }
    \label{fig:SM11}
\end{SCfigure}

\subsubsection*{E.1. Methods}
We detail the mathematical formulation of the biased convergence framework below. 
As it uses the same sliding window technique as the simple contagion framework in the main paper, we follow the same notations as in \S5.2 and focus on the components where the FJ model replaces the SIR model within each time window.

For a retweet network $G(t_n, p) = (V(t_n, p), A(t_n,p))$ generated by aggregating retweets of content type $p$ over the past $n$ months, we first apply the following preprocessing steps to obtain $\hat{G}(t_n, p) = (\hat{V}(t_n, p), \hat{A}(t_n, p))$:

\begin{itemize}

\item To satisfy the FJ convergence criterion (Corollary 22 in \cite{Proskurnikov-2017}), we retain only the union of two node sets and the edges among them: $V_a(t_n, p)$, users aligned with content type $p$ who can reach swayable users, and $V_{sw}(t_n, p)$, the swayable users they can reach. That is, $\hat{V}(t_n, p) = V_a(t_n, p) \cup V_{sw}(t_n, p)$.

\item For the following two reasons related to the FJ model setting, we transpose the network edge direction and normalise the edge weights. That is, we define  $\hat{A}(t_n,p) = (\hat{A}_{i,j}(t_n,p))_{i, j \in \hat{V}(t_n, p)}$, where $\hat{A}_{i,j}(t_n,p) = \frac{A_{j,i}(t_n,p)}{\sum_jA_{j,i}(t_n,p)}$.

\begin{itemize}
\item \textbf{Edge transpose:} 
Recall in the main paper, the retweet network is constructed such that the direction of the edges represents the direction of the information flow, with an edge from user $i$ to $j$ indicating that a tweet from user $i$ captures the interest of user $j$, leading to user $j$ retweeting user $i$. 
In the FJ model, however, edge direction represents the direction of influence, which is opposite to the direction of information flow (i.e., information flowing from user $i$ to user $j$ implies that user $j$ is influenced by user $i$). 
\item \textbf{Weight normalisation:}
The FJ model assumes a network whose weighted adjacency matrix is stochastic, i.e., each row sums to 1 ($\sum_j\hat{A}_{i,j}(t_n,p) = 1$ for all $i \in \hat{V}(t_n, p)$).
\end{itemize}

\end{itemize}

Inspired by Out et al. \cite{Out-2024}, we initialise the FJ model by assigning each node $i \in \hat{V}(t_n, p)$ the following parameters:
\begin{itemize}

\item \textbf{Susceptibility:} The susceptibility of node $i$, denoted as $\lambda_i(p)$, is constant across all time windows, with $\lambda_i(p) \sim U(0, 1)$ if $i \in V_{sw}(t_n, p)$, and $\lambda_i(p) = 0$ otherwise (i.e., $i \in V_a(t_n, p)$). 

\item \textbf{Initial prejudice:} The initial prejudice of node $i$, denoted as $u_i(t_n, p) = x_i^0(t_n, p)$ (with the prejudice assumed equal to the initial opinion), is defined under two scenarios:
\begin{itemize}
\item \textbf{Scenario 1 (memoryless):} $u_i(t_n, p)$ is independent of the time window $t_n$, with $u_i(t_n, p) = 0$ if $i \in V_{sw}(t_n, p)$, and $u_i(t_n, p) = 1$ otherwise (i.e., $i \in V_a(t_n, p)$). 
\item \textbf{Scenario 2 (with memory):}  $u_i(t_n, p)$ is set to the converged opinion from the most recent window, denoted by $x_i^\infty(t_n^-, p)$, if it exists. 
Specifically, 
$$
u_i(t_n, p) = \begin{cases}    1, & if \;  i \in V_a(t_n, p)\\
    x_i^\infty(t_n^-, p), & if \;  i \in V_{sw}(t_n, p) \; and \; x_i^\infty(t_n^-, p) \; exists\\
0, &if \;  i \in V_{sw}(t_n, p) \; and \; x_i^\infty(t_n^-, p)\; does \; not \;exist
\end{cases}
$$
\end{itemize}

\end{itemize}

By Corollary 22 in \cite{Proskurnikov-2017}, the FJ model in our case is asymptotically stable and converges to the unique equilibrium 
\[
\mathbf{x}^\infty(t_n, p)
= (I - \Lambda \hat{A}(t_n, p))^{-1}(I - \Lambda)\mathbf{u}(t_n, p)
\]
where $\Lambda = \mathrm{diag}(\lambda_i)_{i \in \hat{V}(t_n, p)}$, and $I$ is the identity matrix.

Then the estimated follower count increase rate, based on the follower count $f(t_n, i)$ of each user $i \in \hat{V}(t_n, p)$ last recorded before the one-month window, is given by
\[
\hat{r}(t_n, p \mid \delta)
= \frac{\sum_{i \in V_{sw}(t_n, p)} \delta\, f(t_n, i) \cdot \max\{x_i^\infty(t_n, p) - x_i^0(t_n, p), 0\}}
{\sum_{i \in V_a(t_n, p)} f(t_n, i)}
\]
where $\delta$ is a scaling parameter, whose design follows that of the simple contagion framework and remains constant across all time windows and content types.

To account for stochasticity in susceptibility values drawn from a uniform distribution for swayable users, we run 20 realisations for each scenario.
The parameter optimisation follows that of the simple contagion framework, minimising the same quality function $Q$ using the Nelder–Mead numerical method, but involving only a single scaling parameter $\delta$ and no calibration parameter $\mathcal{R}_{t_n}$. 


\end{document}